\magnification=\magstep1

\newbox\SlashedBox
\def\slashed#1{\setbox\SlashedBox=\hbox{#1}
\hbox to 0pt{\hbox to 1\wd\SlashedBox{\hfil/\
hfil}\hss}#1}
\def\hboxtosizeof#1#2{\setbox\SlashedBox=\hbox{#1}
\hbox to 1\wd\SlashedBox{#2}}

\def\mathslashed#1{\setbox\SlashedBox=\hbox{$#1$}
\hbox to 0pt{\hbox to 1\wd\SlashedBox{\hfil/\hfil}\hss}#1}

\def\ifsmall{\iffalse}  
\def\titlepagefont{}  

\def\DefineTeXgraphics{%
\special{ps::[global] /TeXgraphics { } def}}  

\def\today{\ifcase\month\or January\or February\or March\or April\or May
\or June\or July\or August\or September\or October\or November\or
December\fi\space\number\day, \number\year}
\def\eatPrefix19{}
\def\Year{\expandafter\eatPrefix\the\year}
\newcount\hours \newcount\minutes
\def\monthname{\ifcase\month\or
January\or February\or March\or April\or May\or June\or July\or
August\or September\or October\or November\or December\fi}
\def\shortmonthname{\ifcase\month\or
Jan\or Feb\or Mar\or Apr\or May\or Jun\or Jul\or
Aug\or Sep\or Oct\or Nov\or Dec\fi}

\def\TimeStamp{\hours\the\time\divide\hours by60%
\minutes -\the\time\divide\minutes by60\multiply\minutes by60%
\advance\minutes by\the\time%
${\rm \shortmonthname}\cdot\if\day<10{}0\fi\the\day\cdot\the\year%
\qquad\the\hours:\if\minutes<10{}0\fi\the\minutes$}




\newif\ifdraftmode
\newif\ifleftlabels  

\def\nolabels{\def\wrlabeL##1{}\def\eqlabeL##1{}\def\reflabeL##1{}}
\def\writelabels{\def\wrlabeL##1{\leavevmode\vadjust{\rlap{\smash%
{\line{{\escapechar=` \hfill\rlap{\sevenrm\hskip.03in\string##1}}}}}}}%
\def\eqlabeL##1{{\escapechar-1\rlap{\sevenrm\hskip.05in\string##1}}}%
\def\reflabeL##1{\noexpand\rlap{\noexpand\sevenrm[\string##1]}}}
\def\writeleftlabels{\def\wrlabeL##1{\leavevmode\vadjust{\rlap{\smash%
{\line{{\escapechar=` \hfill\rlap{\sevenrm\hskip.03in\string##1}}}}}}}%
\def\eqlabeL##1{{\escapechar-1%
\rlap{\sixrm\hskip.05in\string##1}%
\llap{\sevenrm\string##1\hskip.03in\hbox to \hsize{}}}}%
\def\reflabeL##1{\noexpand\rlap{\noexpand\sevenrm[\string##1]}}}
\nolabels

\newdimen\fullhsize
\newdimen\hstitle
\hstitle=\hsize 
\newdimen\hsbody
\hsbody=\hsize 
\newdimen\hbodyoffset
\hbodyoffset=\hoffset 
\newbox\leftpage
\def\abstract#1{#1}
\def\rotated{\special{ps: landscape}
\magnification=1000  
\baselineskip=14pt
\global\hstitle=9truein\global\hsbody=4.75truein
\global\vsize=7truein\global\voffset=-.31truein
\global\hoffset=-0.54in\global\hbodyoffset=-.54truein
\global\fullhsize=10truein
\def\DefineTeXgraphics{%
\special{ps::[global]
/TeXgraphics {currentpoint translate 0.7 0.7 scale
              -80 0.72 mul -1000 0.72 mul translate} def}}
\let\lr=L
\def\ifsmall{\iftrue}
\def\titlepagefont{\twelvepoint}
\trueseventeenpoint
\def\almostshipout##1{\if L\lr \count1=1
      \global\setbox\leftpage=##1 \global\let\lr=R
   \else \count1=2
      \shipout\vbox{\hbox to\fullhsize{\box\leftpage\hfil##1}}
      \global\let\lr=L\fi}

\output={\ifnum\count0=1 
 \shipout\vbox{\hbox to \fullhsize{\hfill\pagebody\hfill}}\advancepageno
 \else
 \almostshipout{\leftline{\vbox{\pagebody\makefootline}}}\advancepageno
 \fi}

\def\abstract##1{{\leftskip=1.5in\rightskip=1.5in ##1\par}} }

\def\linemessage#1{\immediate\write16{#1}}

\global\newcount\secno \global\secno=0
\global\newcount\appno \global\appno=0
\global\newcount\meqno \global\meqno=1
\global\newcount\subsecno \global\subsecno=0
\global\newcount\figno \global\figno=0

\newif\ifAnyCounterChanged
\let\terminator=\relax
\def\normalize#1{\ifx#1\terminator\let\next=\relax\else%
\if#1i\aftergroup i\else\if#1v\aftergroup v\else\if#1x\aftergroup x%
\else\if#1l\aftergroup l\else\if#1c\aftergroup c\else%
\if#1m\aftergroup m\else%
\if#1I\aftergroup I\else\if#1V\aftergroup V\else\if#1X\aftergroup X%
\else\if#1L\aftergroup L\else\if#1C\aftergroup C\else%
\if#1M\aftergroup M\else\aftergroup#1\fi\fi\fi\fi\fi\fi\fi\fi\fi\fi\fi\fi%
\let\next=\normalize\fi%
\next}
\def\makeNormal#1#2{\def\doNormalDef{\edef#1}\begingroup%
\aftergroup\doNormalDef\aftergroup{\normalize#2\terminator\aftergroup}%
\endgroup}

\def\warnIfChanged#1#2{%
\ifundef#1
\else\begingroup%
\edef\oldDefinitionOfCounter{#1}\edef\newDefinitionOfCounter{#2}%
\ifx\oldDefinitionOfCounter\newDefinitionOfCounter%
\else%
\linemessage{Warning: definition of \noexpand#1 has changed.}%
\global\AnyCounterChangedtrue\fi\endgroup\fi}

\def
\Section#1{\global\advance\secno by1\relax\global\meqno=1%
\global\subsecno=0%
\bigbreak\bigskip
\centerline{\twelvepoint \bf %
\the\secno. #1}%
\par\nobreak\medskip\nobreak}
\def\tagsection#1{%
\warnIfChanged#1{\the\secno}%
\xdef#1{\the\secno}%
\ifWritingAuxFile\immediate\write\auxfile{\noexpand\xdef\noexpand#1{#1}}\fi%
}
\def\section{\Section}
\def\Subsection#1{\global\advance\subsecno by1\relax\medskip %
\leftline{\bf\the\secno.\the\subsecno\ #1}%
\par\nobreak\smallskip\nobreak}
\def\tagsubsection#1{%
\warnIfChanged#1{\the\secno.\the\subsecno}%
\xdef#1{\the\secno.\the\subsecno}%
\ifWritingAuxFile\immediate\write\auxfile{\noexpand\xdef\noexpand#1{#1}}\fi%
}

\def\subsection{\Subsection}

\def\romappno{\uppercase\expandafter{\romannumeral\appno}}
\def\makeNormalizedRomappno{%
\expandafter\makeNormal\expandafter\normalizedromappno%
\expandafter{\romannumeral\appno}%
\edef\normalizedromappno{\uppercase{\normalizedromappno}}}
\def\Appendix#1{\global\advance\appno by1\relax\global\meqno=1\global\secno=0
\bigbreak\bigskip
\centerline{\twelvepoint \bf Appendix %
\romappno. #1}%
\par\nobreak\medskip\nobreak}
\def\tagappendix#1{\makeNormalizedRomappno%
\warnIfChanged#1{\normalizedromappno}%
\xdef#1{\normalizedromappno}%
\ifWritingAuxFile\immediate\write\auxfile{\noexpand\xdef\noexpand#1{#1}}\fi%
}
\def\appendix{\Appendix}

\def\eqn#1{\makeNormalizedRomappno%
\ifnum\secno>0%
  \warnIfChanged#1{\the\secno.\the\meqno}%
  \eqno(\the\secno.\the\meqno)\xdef#1{\the\secno.\the\meqno}%
     \global\advance\meqno by1
\else\ifnum\appno>0%
  \warnIfChanged#1{\normalizedromappno.\the\meqno}%
  \eqno({\rm\romappno}.\the\meqno)%
      \xdef#1{\normalizedromappno.\the\meqno}%
     \global\advance\meqno by1
\else%
  \warnIfChanged#1{\the\meqno}%
  \eqno(\the\meqno)\xdef#1{\the\meqno}%
     \global\advance\meqno by1
\fi\fi%
\eqlabeL#1%
\ifWritingAuxFile\immediate\write\auxfile{\noexpand\xdef\noexpand#1{#1}}\fi%
}
\def\defeqn#1{\makeNormalizedRomappno%
\ifnum\secno>0%
  \warnIfChanged#1{\the\secno.\the\meqno}%
  \xdef#1{\the\secno.\the\meqno}%
     \global\advance\meqno by1
\else\ifnum\appno>0%
  \warnIfChanged#1{\normalizedromappno.\the\meqno}%
  \xdef#1{\normalizedromappno.\the\meqno}%
     \global\advance\meqno by1
\else%
  \warnIfChanged#1{\the\meqno}%
  \xdef#1{\the\meqno}%
     \global\advance\meqno by1
\fi\fi%
\eqlabeL#1%
\ifWritingAuxFile\immediate\write\auxfile{\noexpand\xdef\noexpand#1{#1}}\fi%
}
\def\anoneqn{\makeNormalizedRomappno%
\ifnum\secno>0
  \eqno(\the\secno.\the\meqno)%
     \global\advance\meqno by1
\else\ifnum\appno>0
  \eqno({\rm\normalizedromappno}.\the\meqno)%
     \global\advance\meqno by1
\else
  \eqno(\the\meqno)%
     \global\advance\meqno by1
\fi\fi%
}
\def\mfig#1#2{\global\advance\figno by1%
\relax#1\the\figno%
\warnIfChanged#2{\the\figno}%
\edef#2{\the\figno}%
\reflabeL#2%
\ifWritingAuxFile\immediate\write\auxfile{\noexpand\xdef\noexpand#2{#2}}\fi%
}

\def\fig#1{\mfig{fig.~}#1}

\catcode`@=11 

\font\ninerm=cmr9
\font\eightrm=cmr8
\font\sixrm=cmr6

\def\loadtrueseventeenpoint{
 \font\seventeenrm=cmr10 at 17.28truept
 \font\seventeeni=cmmi10 at 17.28truept
 \font\seventeenbf=cmbx10 at 17.28truept
 \font\seventeenit=cmti10 at 17.28truept
 \font\seventeensl=cmsl10 at 17.28truept
 \font\seventeensy=cmsy10 at 17.28truept
}
\def\loadfourteenpoint{
\font\fourteenrm=cmr10 at 14.4pt
\font\fourteeni=cmmi10 at 14.4pt
\font\fourteenit=cmti10 at 14.4pt
\font\fourteensl=cmsl10 at 14.4pt
\font\fourteensy=cmsy10 at 14.4pt
\font\fourteenbf=cmbx10 at 14.4pt
}
\def\loadtruetwelvepoint{
\font\twelverm=cmr10 at 12truept
\font\twelvei=cmmi10 at 12truept
\font\twelveit=cmti10 at 12truept
\font\twelvesl=cmsl10 at 12truept
\font\twelvesy=cmsy10 at 12truept
\font\twelvebf=cmbx10 at 12truept
}

\font\ninei=cmmi9
\font\eighti=cmmi8
\font\sixi=cmmi6
\skewchar\ninei='177 \skewchar\eighti='177 \skewchar\sixi='177

\font\ninesy=cmsy9
\font\eightsy=cmsy8
\font\sixsy=cmsy6
\skewchar\ninesy='60 \skewchar\eightsy='60 \skewchar\sixsy='60

\font\ninebf=cmbx9
\font\eightbf=cmbx8
\font\sixbf=cmbx6

\font\ninett=cmtt9
\font\eighttt=cmtt8

\hyphenchar\tentt=-1 
\hyphenchar\ninett=-1
\hyphenchar\eighttt=-1

\font\ninesl=cmsl9
\font\eightsl=cmsl8

\font\nineit=cmti9
\font\eightit=cmti8


\newskip\ttglue
\def\tenpoint{\def\rm{\fam0\tenrm}%
  \textfont0=\tenrm \scriptfont0=\sevenrm \scriptscriptfont0=\fiverm
  \textfont1=\teni \scriptfont1=\seveni \scriptscriptfont1=\fivei
  \textfont2=\tensy \scriptfont2=\sevensy \scriptscriptfont2=\fivesy
  \textfont3=\tenex \scriptfont3=\tenex \scriptscriptfont3=\tenex
  \def\it{\fam\itfam\tenit}\textfont\itfam=\tenit
  \def\sl{\fam\slfam\tensl}\textfont\slfam=\tensl
  \def\bf{\fam\bffam\tenbf}\textfont\bffam=\tenbf \scriptfont\bffam=\sevenbf
  \scriptscriptfont\bffam=\fivebf
  \normalbaselineskip=12pt
  \let\sc=\eightrm
  \let\big=\tenbig
  \setbox\strutbox=\hbox{\vrule height8.5pt depth3.5pt width\z@}%
  \normalbaselines\rm}

\def\twelvepoint{\def\rm{\fam0\twelverm}%
  \textfont0=\twelverm \scriptfont0=\ninerm \scriptscriptfont0=\sevenrm
  \textfont1=\twelvei \scriptfont1=\ninei \scriptscriptfont1=\seveni
  \textfont2=\twelvesy \scriptfont2=\ninesy \scriptscriptfont2=\sevensy
  \textfont3=\tenex \scriptfont3=\tenex \scriptscriptfont3=\tenex
  \def\it{\fam\itfam\twelveit}\textfont\itfam=\twelveit
  \def\sl{\fam\slfam\twelvesl}\textfont\slfam=\twelvesl
  \def\bf{\fam\bffam\twelvebf}\textfont\bffam=\twelvebf
  \scriptfont\bffam=\ninebf
  \scriptscriptfont\bffam=\sevenbf
  \normalbaselineskip=12pt
  \let\sc=\eightrm
  \let\big=\tenbig
  \setbox\strutbox=\hbox{\vrule height8.5pt depth3.5pt width\z@}%
  \normalbaselines\rm}

\def\fourteenpoint{\def\rm{\fam0\fourteenrm}%
  \textfont0=\fourteenrm \scriptfont0=\tenrm \scriptscriptfont0=\sevenrm
  \textfont1=\fourteeni \scriptfont1=\teni \scriptscriptfont1=\seveni
  \textfont2=\fourteensy \scriptfont2=\tensy \scriptscriptfont2=\sevensy
  \textfont3=\tenex \scriptfont3=\tenex \scriptscriptfont3=\tenex
  \def\it{\fam\itfam\fourteenit}\textfont\itfam=\fourteenit
  \def\sl{\fam\slfam\fourteensl}\textfont\slfam=\fourteensl
  \def\bf{\fam\bffam\fourteenbf}\textfont\bffam=\fourteenbf%
  \scriptfont\bffam=\tenbf
  \scriptscriptfont\bffam=\sevenbf
  \normalbaselineskip=17pt
  \let\sc=\elevenrm
  \let\big=\tenbig
  \setbox\strutbox=\hbox{\vrule height8.5pt depth3.5pt width\z@}%
  \normalbaselines\rm}

\def\seventeenpoint{\def\rm{\fam0\seventeenrm}%
  \textfont0=\seventeenrm \scriptfont0=\fourteenrm \scriptscriptfont0=\tenrm
  \textfont1=\seventeeni \scriptfont1=\fourteeni \scriptscriptfont1=\teni
  \textfont2=\seventeensy \scriptfont2=\fourteensy \scriptscriptfont2=\tensy
  \textfont3=\tenex \scriptfont3=\tenex \scriptscriptfont3=\tenex
  \def\it{\fam\itfam\seventeenit}\textfont\itfam=\seventeenit
  \def\sl{\fam\slfam\seventeensl}\textfont\slfam=\seventeensl
  \def\bf{\fam\bffam\seventeenbf}\textfont\bffam=\seventeenbf%
  \scriptfont\bffam=\fourteenbf
  \scriptscriptfont\bffam=\twelvebf
  \normalbaselineskip=21pt
  \let\sc=\fourteenrm
  \let\big=\tenbig
  \setbox\strutbox=\hbox{\vrule height 12pt depth 6pt width\z@}%
  \normalbaselines\rm}

\def\ninepoint{\def\rm{\fam0\ninerm}%
  \textfont0=\ninerm \scriptfont0=\sixrm \scriptscriptfont0=\fiverm
  \textfont1=\ninei \scriptfont1=\sixi \scriptscriptfont1=\fivei
  \textfont2=\ninesy \scriptfont2=\sixsy \scriptscriptfont2=\fivesy
  \textfont3=\tenex \scriptfont3=\tenex \scriptscriptfont3=\tenex
  \def\it{\fam\itfam\nineit}\textfont\itfam=\nineit
  \def\sl{\fam\slfam\ninesl}\textfont\slfam=\ninesl
  \def\bf{\fam\bffam\ninebf}\textfont\bffam=\ninebf \scriptfont\bffam=\sixbf
  \scriptscriptfont\bffam=\fivebf
  \normalbaselineskip=11pt
  \let\sc=\sevenrm
  \let\big=\ninebig
  \setbox\strutbox=\hbox{\vrule height8pt depth3pt width\z@}%
  \normalbaselines\rm}

\def\eightpoint{\def\rm{\fam0\eightrm}%
  \textfont0=\eightrm \scriptfont0=\sixrm \scriptscriptfont0=\fiverm%
  \textfont1=\eighti \scriptfont1=\sixi \scriptscriptfont1=\fivei%
  \textfont2=\eightsy \scriptfont2=\sixsy \scriptscriptfont2=\fivesy%
  \textfont3=\tenex \scriptfont3=\tenex \scriptscriptfont3=\tenex%
  \def\it{\fam\itfam\eightit}\textfont\itfam=\eightit%
  \def\sl{\fam\slfam\eightsl}\textfont\slfam=\eightsl%
  \def\bf{\fam\bffam\eightbf}\textfont\bffam=\eightbf \scriptfont\bffam=\sixbf%
  \scriptscriptfont\bffam=\fivebf%
  \normalbaselineskip=9pt%
  \let\sc=\sixrm%
  \let\big=\eightbig%
  \setbox\strutbox=\hbox{\vrule height7pt depth2pt width\z@}%
  \normalbaselines\rm}

\def\tenbig#1{{\hbox{$\left#1\vbox to8.5pt{}\right.\n@space$}}}
\def\ninebig#1{{\hbox{$\textfont0=\tenrm\textfont2=\tensy
  \left#1\vbox to7.25pt{}\right.\n@space$}}}
\def\eightbig#1{{\hbox{$\textfont0=\ninerm\textfont2=\ninesy
  \left#1\vbox to6.5pt{}\right.\n@space$}}}

\def\footnote#1{\edef\@sf{\spacefactor\the\spacefactor}#1\@sf
      \insert\footins\bgroup\eightpoint
      \interlinepenalty100 \let\par=\endgraf
        \leftskip=\z@skip \rightskip=\z@skip
        \splittopskip=10pt plus 1pt minus 1pt \floatingpenalty=20000
        \smallskip\item{#1}\bgroup\strut\aftergroup\@foot\let\next}
\skip\footins=12pt plus 2pt minus 4pt 
\dimen\footins=30pc 

\newinsert\margin
\dimen\margin=\maxdimen

\loadtruetwelvepoint 
\loadtrueseventeenpoint
\catcode`\@=\active
\catcode`@=12  
\catcode`\"=\active

\def\eatOne#1{}
\def\ifundef#1{\expandafter\ifx%
\csname\expandafter\eatOne\string#1\endcsname\relax}
\def\notTrue{\iffalse}\def\isTrue{\iftrue}
\def\ifdef#1{{\ifundef#1%
\aftergroup\notTrue\else\aftergroup\isTrue\fi}}
\def\use#1{\ifundef#1\linemessage{Warning: \string#1 is undefined.}%
{\tt \string#1}\else#1\fi}


\global\newcount\refno \global\refno=1
\newwrite\rfile
\newlinechar=`\^^J
\def\ref#1#2{\the\refno\nref#1{#2}}
\def\nref#1#2{\xdef#1{\the\refno}%
\ifnum\refno=1\immediate\openout\rfile=refs.tmp\fi%
\immediate\write\rfile{\noexpand\item{[\noexpand#1]\ }#2.}%
\global\advance\refno by1}
\def\lref#1#2{\the\refno\xdef#1{\the\refno}%
\ifnum\refno=1\immediate\openout\rfile=refs.tmp\fi%
\immediate\write\rfile{\noexpand\item{[\noexpand#1]\ }#2\semi}%
\global\advance\refno by1}
\def\cref#1{\immediate\write\rfile{#1\semi}}

\def\semi{;\hfil\noexpand\break}

\def\inputAuxIfPresent#1{\immediate\openin1=#1
\ifeof1\message{No file \auxfileName; I'll create one.
}\else\closein1\relax\input\auxfileName\fi%
}

\newif\ifWritingAuxFile
\newwrite\auxfile
\def\SetUpAuxFile{%
\xdef\auxfileName{\jobname.aux}%
\inputAuxIfPresent{\auxfileName}%
\WritingAuxFiletrue%
\immediate\openout\auxfile=\auxfileName}


\def\bye{\par\vfill\supereject%
\ifAnyCounterChanged\linemessage{
Some counters have changed.  Re-run tex to fix them up.}\fi%
\end}

\SetUpAuxFile

\catcode`@=11  
\def\meqalign#1{\,\vcenter{\openup1\jot\m@th
   \ialign{\strut\hfil$\displaystyle{##}$ && $\displaystyle{{}##}$\hfil
             \crcr#1\crcr}}\,}
\catcode`@=12  


\baselineskip 15pt
\overfullrule 0.5pt

%
\newread\epsffilein    
\newif\ifepsffileok    
\newif\ifepsfbbfound   
\newif\ifepsfverbose   
\newdimen\epsfxsize    
\newdimen\epsfysize    
\newdimen\epsftsize    
\newdimen\epsfrsize    
\newdimen\epsftmp      
\newdimen\pspoints     
\pspoints=1bp          
\epsfxsize=0pt         
\epsfysize=0pt         
\def\epsfbox#1{\global\def\epsfllx{72}\global\def\epsflly{72}%
   \global\def\epsfurx{540}\global\def\epsfury{720}%
   \def\lbracket{[}\def\testit{#1}\ifx\testit\lbracket
   \let\next=\epsfgetlitbb\else\let\next=\epsfnormal\fi\next{#1}}%
\def\epsfgetlitbb#1#2 #3 #4 #5]#6{\epsfgrab #2 #3 #4 #5 .\\%
   \epsfsetgraph{#6}}%
\def\epsfnormal#1{\epsfgetbb{#1}\epsfsetgraph{#1}}%
\def\epsfgetbb#1{%
%
%
\openin\epsffilein=#1
\ifeof\epsffilein\errmessage{I couldn't open #1, will ignore it}\else
%
%
   {\epsffileoktrue \chardef\other=12
    \def\do##1{\catcode`##1=\other}\dospecials \catcode`\ =10
    \loop
       \read\epsffilein to \epsffileline
       \ifeof\epsffilein\epsffileokfalse\else
%
%
          \expandafter\epsfaux\epsffileline:. \\%
       \fi
   \ifepsffileok\repeat
   \ifepsfbbfound\else
    \ifepsfverbose\message{No bounding box comment in #1; using defaults}\fi\fi
   }\closein\epsffilein\fi}%
%
%
\def\epsfclipstring{}
\def\epsfsetgraph#1{%
   \epsfrsize=\epsfury\pspoints
   \advance\epsfrsize by-\epsflly\pspoints
   \epsftsize=\epsfurx\pspoints
   \advance\epsftsize by-\epsfllx\pspoints
%
%
   \epsfxsize\epsfsize\epsftsize\epsfrsize
   \ifnum\epsfxsize=0 \ifnum\epsfysize=0
      \epsfxsize=\epsftsize \epsfysize=\epsfrsize
      \epsfrsize=0pt
%
%
     \else\epsftmp=\epsftsize \divide\epsftmp\epsfrsize
       \epsfxsize=\epsfysize \multiply\epsfxsize\epsftmp
       \multiply\epsftmp\epsfrsize \advance\epsftsize-\epsftmp
       \epsftmp=\epsfysize
       \loop \advance\epsftsize\epsftsize \divide\epsftmp 2
       \ifnum\epsftmp>0
          \ifnum\epsftsize<\epsfrsize\else
             \advance\epsftsize-\epsfrsize \advance\epsfxsize\epsftmp \fi
       \repeat
       \epsfrsize=0pt
     \fi
   \else \ifnum\epsfysize=0
     \epsftmp=\epsfrsize \divide\epsftmp\epsftsize
     \epsfysize=\epsfxsize \multiply\epsfysize\epsftmp
     \multiply\epsftmp\epsftsize \advance\epsfrsize-\epsftmp
     \epsftmp=\epsfxsize
     \loop \advance\epsfrsize\epsfrsize \divide\epsftmp 2
     \ifnum\epsftmp>0
        \ifnum\epsfrsize<\epsftsize\else
           \advance\epsfrsize-\epsftsize \advance\epsfysize\epsftmp \fi
     \repeat
     \epsfrsize=0pt
    \else
     \epsfrsize=\epsfysize
    \fi
   \fi
%
%
   \ifepsfverbose\message{#1: width=\the\epsfxsize, height=\the\epsfysize}\fi
   \epsftmp=10\epsfxsize \divide\epsftmp\pspoints
   \vbox to\epsfysize{\vfil\hbox to\epsfxsize{%
      \ifnum\epsfrsize=0\relax
        \includegraphics{#1}%
      \else
        \epsfrsize=10\epsfysize \divide\epsfrsize\pspoints
        \includegraphics{#1}%
      \fi
      \hfil}}%
\global\epsfxsize=0pt\global\epsfysize=0pt}%
%
%
{\catcode`\%=12 \global\let\epsfpercent=
%
%
\long\def\epsfaux#1#2:#3\\{\ifx#1\epsfpercent
   \def\testit{#2}\ifx\testit\epsfbblit
      \epsfgrab #3 . . . \\%
      \epsffileokfalse
      \global\epsfbbfoundtrue
   \fi\else\ifx#1\par\else\epsffileokfalse\fi\fi}%
%
%
\def\epsfempty{}%
\def\epsfgrab #1 #2 #3 #4 #5\\{%
\global\def\epsfllx{#1}\ifx\epsfllx\epsfempty
      \epsfgrab #2 #3 #4 #5 .\\\else
   \global\def\epsflly{#2}%
   \global\def\epsfurx{#3}\global\def\epsfury{#4}\fi}%
%
%
\def\epsfsize#1#2{\epsfxsize}
%
%



\def\ref#1#2{\nref#1{#2}}
\overfullrule 0pt
\hfuzz 40pt
\hsize 6. truein
\vsize 8.5 truein

\loadfourteenpoint
\newcount\eqncount
\newcount\sectcount
\eqncount=0
\sectcount=0
\def\secta{\global\advance\sectcount by1
\eqncount=0}

\def\equn{
\global\advance\eqncount by1
\eqno{(\the\sectcount.\the\eqncount)}        }
\def\put#1{\global\edef#1{(\the\sectcount.\the\eqncount)}     }

\def\section#1{\global\advance\secno by1\relax\global\meqno=1%
\global\subsecno=0%
\bigbreak\bigskip
\noindent{\twelvepoint \bf %
\the\secno. #1}%
\par\nobreak\medskip\nobreak}

\noindent
IEM-FT-96-133 \hfill DESY 96-107\break
hep-ph/9606316 \hfill  \break

\vskip 1.5 cm

\baselineskip 15 pt
\centerline{\bf THEORETICAL HIGGS MASS BOUNDS}
\vskip .1 cm
\centerline{\bf IN THE STANDARD MODEL}
\vskip .1 cm
\centerline{\bf AND SUPERSYMMETRIC EXTENSIONS
\footnote{${}^*$}%
{Lectures presented at the XXIV ITEP Winter School, Snegiri (Russia)
February 1996.} }

\vskip 3. cm
\centerline{Jose Ram\'on Espinosa} 
\centerline{\it 
DESY Theory Group} \centerline{\it Deutsches Elektronen Synchrotron DESY}
\centerline{\it 22603 Hamburg, Germany}


\vskip 2.5 truecm

\vskip 1. cm
{ \narrower\smallskip
\centerline{\bf Abstract}
\vskip .1 cm
These lectures provide a very basic introduction to different theoretical 
limits on the mass of Higgs scalars. Particular attention is devoted to 
the pure Standard Model and its Minimal Supersymmetric extension 
(MSSM).
\smallskip}


\vfill
\break


\baselineskip 12 pt
\centerline{\bf THEORETICAL HIGGS MASS BOUNDS}
\vskip .1 cm
\centerline{\bf IN THE STANDARD MODEL}
\vskip .1 cm
\centerline{\bf AND SUPERSYMMETRIC EXTENSIONS}
\vskip .6 cm
\centerline{ Jose Ram\'on Espinosa}
\centerline{\it DESY Theory Group}
\centerline{\it Deutsches Elektronen Synchrotron DESY}
\centerline{\it 22603 Hamburg, Germany}


\baselineskip12pt

\vskip .6 cm
{ \narrower\smallskip
\centerline{\bf Abstract}
\ninerm\baselineskip 10 pt
These lectures provide a very basic introduction to different theoretical 
limits on the mass of Higgs scalars. Particular attention is devoted to 
the pure Standard Model and its Minimal Supersymmetric extension 
(MSSM).
\smallskip}

\vskip 0.3 truecm


\baselineskip 12 pt
\section{Introduction and Overview}
\tagsection\IntroductionSection

Uncovering the elusive mechanism of electroweak symmetry breaking is one of 
the main goals of present and future accelerators. In these lectures I 
will concentrate on the most popular and simplest of all proposed mechanisms 
for that breaking. It makes use of a sector of 
fundamental scalar particles and
the breaking of $SU(2)\times U(1)$ is achieved spontaneously by the vacuum
expectation values (vevs) of some of these scalar fields, which have in 
general non-vanishing electroweak charges.

A generic prediction of these models is the existence of physical scalar
particles, the Higgs bosons, remnant of the electroweak breaking searched
for in accelerator experiments. 
The aim of these lectures is to give an introduction to some theoretical 
guide available for that search in the form of Higgs mass bounds. The first 
section starts with some general statements that can be made with the 
only assumption that the Higgs sector is weakly interacting. Particularly 
relevant examples of models, like the pure Standard Model or its minimal 
Supersymmetric extensions are then put in a clearer perspective by 
contrast with the general case.

The precise computation of an upper bound on the mass of the lightest Higgs
boson in the Minimal Supersymmetric Standard Model is the topic of Section~3,
while 4 and 5 are devoted to an equally precise computation of a lower 
bound on the Standard Model Higgs mass, from studies of the effective 
potential structure. These bounds lie in a mass region especially 
appealing for Higgs searches in the near future and are thus very relevant. 
Some implications that would follow from a Higgs discovery in such 
mass region are considered in Section~6.

\section{Limits from spontaneously broken symmetries}
\tagsection\Spontex

\noindent
{\it \Spontex.1 General Sum Rules}
\vskip .1 cm

Let $\{\phi_i\}$ be the set of Hermitian spinless fields in the theory
with a potential $V(\phi_i)$ not necessarily polynomial, but invariant
under some continuous symmetry G (global or local):
$$
\phi_i \rightarrow \phi_i +\epsilon_\alpha \theta^\alpha_{ij}\phi_j,
\eqn\gtrans
$$
where the $\theta^\alpha$'s are the generators of G, antisymmetric in our 
Hermitian basis.

We are interested in the case of spontaneously broken G, so that the 
minimum of $V(\phi_i)$ occurs at $\phi_i={\tilde v}_i$ with 
$(\theta^\alpha{\tilde v})_i\neq 0$ for some of the $\alpha$'s. Furthermore,
we will be interested in the scalar spectrum of states in the broken minimum
${\tilde v}_i$. The first derivative of $V$ with respect to the fields 
$\phi_i$ will be zero at the minimum by definition. The second derivative,
$$
M^2_{ij}={\partial^2 V 
\over\partial\phi_i\partial\phi_j}\biggr|_{\phi={\tilde v}},
$$
gives the scalar mass matrix while higher order derivatives give 
scalar self-couplings
$$
e_{ijk}={\partial^3 V 
\over\partial\phi_i\partial \phi_j\partial\phi_k}\biggr|_{\phi={\tilde 
v}},\;\;\;\; f_{ijkl}={\partial^4 V 
\over\partial\phi_i\partial\phi_j\partial
\phi_k\partial\phi_l}\biggr|_{\phi={\tilde v}},\cdots
$$
Our starting point is the invariance of the scalar
potential under the transformation (\use\gtrans):
$$
V(\phi_i +\epsilon_\alpha \theta^\alpha_{ij}\phi_j)=
V(\phi_i)+\epsilon_\alpha {\partial V\over\partial 
\phi_i}\theta_{ik}^\alpha\phi_k +\cdots=V(\phi_i)
\anoneqn
$$
from which we obtain the identities
$$
W^\alpha(\phi_i)\equiv {\partial V\over\partial 
\phi_i}\theta_{ik}^\alpha\phi_k\equiv 0.
\eqn\wal
$$
Taking derivatives of these functions $W^\alpha$ with respect to the 
$\phi_i$ fields and evaluating them at the minimum, relations among 
different parameters of the potential can be obtained.
The first derivatives give
$$
{\partial W^\alpha\over\partial\phi_j}\equiv 0 \Rightarrow 
0=M^2_{ij}(\theta^\alpha{\tilde v})_i,
\eqn\first
$$
which is just the familiar Goldstone's theorem: for every generator that 
does not annihilate the vacuum, $M^2$ has a zero eigenvalue.
We note at this point that ${\tilde v}_i$ can be written in general as a sum
of a G-singlet $s_i$ and a non singlet $v_i$. As $\theta^\alpha {\tilde v}=
\theta^\alpha v +\theta^\alpha s =\theta^\alpha v$ we will drop the tilde 
in the following. 
The second derivatives of (\use\wal) give a relation between masses and cubic
couplings:
$$
{\partial^2 W^\alpha\over\partial\phi_j\partial\phi_k}\equiv 0 \Rightarrow 
0=e_{ijk}(\theta^\alpha v)_i + [M^2,\theta^\alpha]_{jk},
\eqn\second
$$
while the third derivatives relate quartic and cubic couplings:
$$
{\partial^3 W^\alpha\over\partial\phi_j\partial\phi_k\partial\phi_l}\equiv 0 
\Rightarrow 0=f_{ijkl}(\theta^\alpha v)_i 
+ e_{ijk}\theta^\alpha_{il}
+ e_{ikl}\theta^\alpha_{ij}
+ e_{ilj}\theta^\alpha_{ik},
\eqn\third
$$
(sum over repeated indices is assumed).
Now, it is simple to eliminate cubic couplings between (\use\second) and 
(\use\third) to obtain a sum rule between masses and (adimensional) quartic 
couplings
$$
\eqalign{
M^2_{ij}[(\theta^\alpha\theta^\beta v)_i(\theta^\gamma\theta^\delta v)_j +
(\theta^\alpha\theta^\gamma v)_i(\theta^\beta\theta^\delta v)_j +
(\theta^\alpha\theta^\delta v)_i(\theta^\beta\theta^\gamma v)_j]&\cr
 =
f_{ijkl}(\theta^\alpha 
v )_i(\theta^\beta v )_j(\theta^\gamma v )_k(\theta^\delta v)_l.& }
\eqn\sumrule $$
This is the central relation of this subsection. It implies that 
knowledge about the dimensionless scalar quartic couplings can be used to 
relate some scalar mass to the G breaking scale set by $v_i$, which is the
only other dimensionful parameter entering (\use\sumrule). We will have 
occasion to see similar mass sum rules later on. Now it would be 
interesting to particularize (\use\sumrule) to the electroweak gauge group
breaking and this example will provide a clearer picture of the relevance of 
(\use\sumrule).
\vskip .5cm  

\noindent {\it 2.1 APPLICATION: $G=SU(2)_L\times U(1)_Y\rightarrow 
U(1)_{em}$}
\vskip .3cm 

There are three broken generators that we choose to be $\theta^\alpha=\{
T_3,T_+,T_-\}$. Corresponding to these generators there will appear three
Goldstone bosons (to be eaten by the gauge bosons). Those are given by the
(complex) vectors
$$
z_i={1\over N}(T_3 v)_i,\;\;\;\;w_i={1\over N'}(T_-v)_i,
$$
where $N,N'$ are normalization constants related to the gauge boson 
masses by $M_Z=g N/\cos\theta_W$ and $M_W=gN'/\sqrt{2}$.

Conservation of electric charge implies there are only three 
non-trivial mass sum rules to be derived from (\use\sumrule)
corresponding to the choices of  
$\{\theta^\alpha,\theta^\beta,\theta^\gamma,\theta^\delta\}$ with charges 
adding up to zero. The scalar mass matrix will have block diagonal form 
breaking up in different submatrices for the differently charged scalar 
particles. Using a mass eigenstate basis, $M^2_{ij}=M^2_i\delta_{ij}$
the following mass sum rules are obtained
$$
\eqalign{
f_{Z}&=3\sum_j[M^{(0)}_j]^2A_j^2,\cr
f_{ZW}&=
\sum_j\left\{[M^{(0)}_j]^2A_jB_j+2[M^{(+)}_j]^2|C_j|^2\right\}
,\cr
f_{W}&=
\sum_j\left\{2[M^{(0)}_j]^2B_j^2+[M^{(++)}_j]^2|D_j|^2\right\}
 },
\eqn\sumew
$$
where
$f_{Z}=f_{ijkl}z_iz_jz_kz_l,
f_{ZW}=f_{ijkl}z_iz_jw_kw_l^*,
f_{W}=f_{ijkl}w_iw_j^*w_kw_l^*,$
are the quartic couplings of Goldstone bosons and
$$
A_j={1\over N^2}(T_3^2v)_j,\;\;\;\;\;\;\;\;\;\;\;
B_j={1\over N'^2}[(T(T+1)-T_3^2)v]_j.
$$
In (\use\sumew), superindices in mass matrices indicate the 
charge of the states.

\noindent{\bf Exercise.} Obtain the corresponding expressions for $C_j$ and
$D_j$. Then prove the following relations
$$
\eqalign{
\sum_j A_jv_j&=\sum_j B_jv_j=1,\;\;\;\;A_jB_j>0,\cr
\sum_jA_jB_j&=\sum_j|C_j|^2+\sqrt{2}{G_F\over\rho^2},
}
\eqn\ex
$$
where $G_F$ is Fermi's constant, $(\sqrt{2}G_F)^{-1}=(246 GeV)^2$, and 
$\rho=M_W/(M_Z\cos\theta_W)$.

It is easy to transform these sum rules into useful mass inequalities.
Consider the first equality in (\use\sumew). If every (neutral) mass 
eigenvalue $M_j^{(0)}$ is substituted by the lowest one $M_{min}^{(0)}$
it follows that
$$
3[M_{min}^{(0)}]^2\sum_jA_j^2\leq f_{Z}.
$$
Furthermore, using the relations (\use\ex) and the Schwartz inequality, 
it follows
$$
\sum_jA_j^2\geq {(\sum_iA_iv_i)^2\over \sum_kv_k^2}={1\over 
\sum_kv_k^2}\equiv {1\over v^2}.
$$
Then one obtains
$$
[M_{min}^{(0)}]^2\leq{1\over 3}f_Zv^2
\leq {1\over 3}f_Z(\sqrt{2}G_F)^{-1},
$$ (note 
that possibly 
large singlet vacuum expectation values were dropped from the $v_k$'s).
By similar methods the following inequalities can be obtained from 
(\use\sumew):
$$
[M_{min}^{(0)}]^2\leq\left\{
{f_{Z}\over3\sqrt{2}G_F},{f_{ZW}\over
\sqrt{2}G_F}\rho^2,{f_{W}\over
2\sqrt{2}G_F} 
\right\}.
\eqn\inequ
$$
We can see explicitly here that provided the quartic couplings $f$ are 
weak one expects always a scalar with mass fixed by the electroweak
scale. This holds irrespective of how complicated the Higgs sector
is. In the case of the minimal Standard Model, with electroweak
breaking described by a single Higgs doublet, the above inequalities
are the same. In fact, the inequalities are saturated and one finds the
familiar relation between the mass and quartic Higgs coupling of the
Higgs boson, $M_h^2=\lambda v^2$. In the general case there will always
appear some scalar state whose mass cannot be made much bigger than
the electroweak scale without making large some dimensionless coupling.
It is natural to refer to that state as the 'true' Higgs boson. In 
general, the 
masses of other scalar states can instead be made heavy by choosing large
values of the mass parameters in the lagrangian.
\vskip .2cm

\noindent
{\it \Spontex.2 Quartic Polynomial Potentials}
\vskip .1 cm

One remarkable property of the sum rules (\use\sumrule) obtained in the
previous subsection is that the only assumption on the form of the potential
was the requirement of invariance under the action of G. Then, these sum
rules can be obtained even if the potential is non-polynomial (e.g. in
supergravity, or in potentials including radiative corrections).
However, it will prove illustrative to consider a somewhat less general
situation to gain some intuitive understanding on the relation between
the spontaneous breaking of some continuous symmetry and the mass limits
implied by it. In this subsection we will then concentrate in cases where 
the potential is indeed polynomial and at most quartic in the scalar 
fields. From the spontaneous breaking of the symmetry G of the potential we 
will obtain sum rules different to those derived in the previous subsection.

Let us look more closely to the potential along the direction of the
symmetry breaking. For this purpose use a mass eigenstate basis $\{\phi_A\}$.
We can decompose every $\phi_A$ in a G-singlet part, $s_A$, and a non-singlet
part $h_A$. Fixing all singlet parts to their vevs the potential for 
the remaining $h_A$-fields is trivially invariant under G. In particular, this
implies that no linear terms in the $h_A$-fields will be allowed.
Now consider the breaking direction in $h_A$-space. Call $\{n_A\}$ a unit 
vector pointing in the direction of the breaking: $n_A=\langle h_A\rangle/v$
and $h$ the normalized field along that direction: $h=\sum_An_Ah_A$. All the
G-breaking is then given by the vev of the state $h$, while all other
orthogonal states in $h_A$-space will have zero vev.
The potential along $h$ can then be written in the form
$$
V(h)=V(0)+{1\over2}\mu^2h^2+{1\over3}\sigma h^3+{1\over 4}\lambda_h h^4,
\eqn\potbd
$$
The value of $h$ at the minimum, $\langle h\rangle=v$, can be related to 
the 
parameters in the potential via the minimization conditions as usual
$$
{\partial V\over \partial h}=0\Rightarrow \mu^2+\sigma v +\lambda_h v^2=0.
\eqn\mini
$$
The second derivative at the minimum can be written as
$$
{\partial^2V\over \partial h^2}=\sum_{A,B}n_An_B{\partial^2V\over\partial 
\phi_A\partial\phi_B}=\sum_A M_A^2n_A^2=\mu^2+2\sigma v +3 \lambda_h v^2
$$
or, using (\use\mini)
$$
\sum_A M_A^2n_A^2=\sigma v +2\lambda_h v^2.
\eqn\sumd
$$
This sum rule does not seem to be particularly useful because the right hand 
side contains some unknown mass parameter $\sigma$. However it is simple to 
see that the quantity $\sigma v$ is always negative in the true minimum
of the potential (\use\potbd) [the degeneracy of the two minima of 
(\use\potbd) at $\sigma=0$ is lifted by the term $\sigma h^3$. The true
minimum will then correspond to $\sigma v^3<0$].
As a result, $\sigma v$ can be dropped in (\use\sumd) to obtain the 
mass limit 
$$
\sum_AM_A^2n_A^2\leq 2\lambda_h v^2 \Rightarrow [M_{min}]^2\leq 
2\lambda_h v^2.
\eqn\ineqd
$$
The last mass inequality follows the same line of Sect. 2.1 and uses 
$\sum n_A^2=1$.
Note that now, the quartic coupling $\lambda_h$ is not necessarily related
to the Goldstone couplings that appear in the mass inequalities of 
(\use\inequ) implying that in general, the mass bound (\use\ineqd) 
is different from those.

We can improve the previous derivation if the symmetry group $G$ contains an
$SU(2)$ subgroup (this apply in particular to the electroweak gauge 
group and is the reason why we concentrate on it. One can construct a 
similar line of derivation for any $U(1)$ subgroup of $G$). In that case,
the Hermitian scalar fields $\phi_j$ will belong to some $SU(2)$ multiplet
of dimension $2T_j+1$. We can define the following parity transformation
$$
P:\;\;\;\phi_j\rightarrow (-1)^{2T_j}\phi_j.
$$
This transformation can indeed be defined for all the fields in the theory
and also for products of any number of fields. Call even the
fields (or product of fields) invariant under $P$ and odd  the fields
that change sign under $P$. In particular, $SU(2)$ singlets, triplets, etc
are even while doublets, quadruplets, etc are odd. 

By fixing all even fields to their vevs we can obtain the potential for
odd fields only. In the odd-field space we then define the normalized field 
$$
\varphi={1\over v_o}\sum_{odd} v_i\phi_i,
$$
along the direction of the breaking. Again, all combinations of odd fields
orthogonal to $\varphi$ have zero vev and the potential $V(\varphi)$ must
be $P$-invariant. Then, not only linear terms in $\varphi$ are forbidden, 
but also cubic ones:
$$
V(\varphi)=V(0)-{1\over 2}m^2\varphi^2+{1\over 8}\lambda_{\varphi}\varphi^4.
$$
If $\varphi$ were a mass eigenvalue it would have mass 
$M_{\varphi}^2=\lambda_{\varphi}v_o^2$. In the general case it is clear 
that some eigenvalue will have mass below that value and the following
inequality follows:
$$
[M_{min}]^2\leq \lambda_{\varphi} v_o^2.
\eqn\ineqt
$$
Note that, in general, both the quartic coupling and the vev appearing in 
this mass bound will be different from the ones derived previously.
\vskip .5cm  

\noindent {\it 2.2 APPLICATION: Lightest Higgs in SUSY extended models}
\vskip .3cm 

It is straightforward to apply the mass bound (\use\ineqt) to derive 
a tree level upper bound on the mass of the lightest neutral Higgs boson in 
general supersymmetric models. 
The quartic coupling for $\varphi$ is obtained from two sources: F and D 
terms. The only terms in the superpotential than can contribute to 
$\lambda_{\varphi}$ through F-terms will have the form
$W_3\sim h \Phi_e\Phi_o\Phi_o$ (the subindices indicate whether
the chiral superfields $\Phi$ are even or odd under $P$). The contribution 
of such terms to $V(\varphi)$ will be then
$$
\delta_F V(\varphi)={1\over 4}\sum_i h_i^2c_i^2 \varphi^4,
\eqn\vf
$$
where the dimensionless Yukawa couplings $h_i$'s are in general accompanied 
by some rotation angles coming from the projection of the odd fields into the
$\varphi$ direction: $c_i=v_i/v_o$.
The contribution of D-terms depends on the gauge group. Assuming the minimal
$SU(2)_L\times U(1)_Y$ it is easy to obtain
$$
\delta_D V(\varphi)={1\over 8}(g^2+g'^2)\left[\sum_{i,odd} Y_i 
c_i^2\right]^2 \varphi^4. \eqn\vd
$$
Adding (\use\vf) and (\use\vd), eq.~(\use\ineqt) gives the following mass 
limit 
$$
M_{h^0}^2\leq (g^2+g'^2)\left[\sum_{i,odd} Y_i c_i^2\right]^2v_o^2+
2\sum_i h_i^2c_i^2v_o^2.
\eqn\susyb
$$

In softly broken SUSY models, the details of
the supersymmetry breaking do not affect the quartic couplings (at tree
level) so that the scale of the mass bound (\use\ineqt) will be fixed only
by the Fermi constant. As we have seen, this is not a particular 
feature of supersymmetric models but holds with complete generality.
The Minimal Supersymmetric Standard Model has the particularity that 
only gauge couplings appear in (\use\susyb). As these are measured 
experimentally, a well defined tree level mass bound results. 
In general, numerical values for the bound (\use\susyb) can be obtained 
only by setting upper limits on the (unknown) $h_i$ couplings, e.g. by 
triviality arguments, requiring that they remain perturbative up to some 
large energy scale.

\noindent{\bf Exercise.} For spontaneous breaking of the electroweak 
gauge group: a) if driven only by $SU(2)$ doublets, show that the bounds
(\use\inequ,\use\ineqd,\use\ineqt) coincide; b) if the scalar potential 
has a $SU(2)$ custodial symmetry after the breaking, show that the bounds 
(\use\inequ) are equal.
\vskip .2cm

\noindent
{\it \Spontex.3 Interplay between different bounds and the decoupling limit}
\vskip .1 cm

All mass bounds derived so far have in common the following 
property: they follow from a sum rule of the form ($a$ 
simply counts different bounds)
$$
\langle\varphi_a|M^2|\phi_a\rangle=\lambda_av_a^2,
\eqn\vect
$$
where $\varphi_a$ and $\phi_a$ are certain scalar fields [that can in 
general be normalized such that $\langle \varphi_a|\phi_a\rangle=1$] , 
$M^2$ is the scalar mass matrix, $\lambda_a$ is some scalar quartic coupling 
and $v_a$ is of the order of the breaking scale. We will call (\use\vect)
the `vector-form' of the bounds.
From them,  the usual 
`scalar-form' of the mass bounds are derived straightforwardly by repeated 
use of $I=\sum_A|A\rangle 
\langle A|$, where the $|A\rangle$ are mass eigenvalues: 
$$
\langle\varphi_a|M^2|\phi_a\rangle=\lambda_av_a^2=\sum_{A}
\langle\varphi_a|A\rangle M_A^2\langle A|\phi_a\rangle\geq M_h^2\sum_A
\langle\varphi_a|A\rangle\langle A|\phi_a\rangle=M_h^2.
$$
In table 1 we present all five mass bounds obtained previously in  
vector and scalar forms. The first column gives the vectors entering
(\use\vect). When only one vector is shown it is assumed $\varphi=\phi$.
Note that only the second bound has different vectors. 

\noindent{\bf Exercise.} If $\langle \varphi |M^2|\varphi\rangle \leq 
\lambda_{\varphi}v_{\varphi}^2$ and 
$\langle \phi |M^2|\phi\rangle \leq \lambda_{\phi}v_{\phi}^2$
with $|\varphi\rangle \neq \kappa|\phi\rangle$, then 
$\langle \varphi |M^2|\phi\rangle$ has also a bound of the form 
$\lambda v^2$. a) find $\lambda$ and $v^2$; b) apply this to (\use\inequ).

\vskip -.2 truecm

$$
\vbox{\offinterlineskip
\halign{ &\vrule# & \strut\quad\hfil#\quad\cr
\omit&\multispan7\hrulefill&\cr
height2pt&\omit&&\omit&&\omit&&\omit&\cr
& \vtop to 0pt{\hbox{\vbox to 7pt{}}\hbox{
$|\phi_a\rangle,\,\;\;|\varphi_a\rangle$ }\vss}\hfil &
& \vtop to 0pt{\hbox{\vbox to 7pt{}}\hbox{$\lambda_a$ }\vss}\hfil
&& \vtop to 0pt{\hbox{\vbox to 7pt{}}\hbox{$v_a^2$ }\vss}\hfil
&& \vtop to 0pt{\hbox{\vbox to 7pt{}}\hbox{
$M_h^2\leq $ }\vss}\hfil
&&\cr
height2pt&\omit&&\omit&&\omit&&\omit&\cr
height2pt&\omit&&\omit&&\omit&&\omit&\cr
&  && && && &\cr
\omit&\multispan7\hrulefill&\cr
height2pt&\omit&&\omit&&\omit&&\omit&\cr
\omit&\multispan7\hrulefill&\cr
height2pt&\omit&&\omit&&\omit&&\omit&\cr
& \vtop to 0pt{\hbox{\vbox to 7pt{}}\hbox{
$\phi_Z\sim\sum_it^2_{3i}v_i\phi_i$ }\vss}\hfil
&& \vtop to 0pt{\hbox{\vbox to 7pt{}}\hbox{
${1\over 24}\lambda_{00}(G^0)^4$ }\vss}\hfil
&& \vtop to 0pt{\hbox{\vbox to 7pt{}}\hbox{ $\sum_iv_i^2$ }\vss}\hfil
&& \vtop to 0pt{\hbox{\vbox to 7pt{}}\hbox{
${1\over 3}\lambda_{00}v^2 $ }\vss}\hfil
&&\cr
height2pt&\omit&&\omit&&\omit&&\omit&\cr
height2pt&\omit&&\omit&&\omit&&\omit&\cr
&  &&    && && &\cr
height2pt&\omit&&\omit&&\omit&&\omit&\cr
\omit&\multispan7\hrulefill&\cr
height2pt&\omit&&\omit&&\omit&&\omit&\cr
& \vtop to 0pt{\hbox{\vbox to 7pt{}}\hbox{  
$\varphi\sim \phi_Z,\,\,\phi\sim\phi_W$  }\vss}\hfil
&
& \vtop to 0pt{\hbox{\vbox to 7pt{}}\hbox{
${1\over 2}\lambda_{0c}|G^0G^+|^2$ }\vss}\hfil
&& \vtop to 0pt{\hbox{\vbox to 7pt{}}\hbox{
 ${\rho^2\over\sqrt{2}}G_F^{-1}$ }\vss}\hfil
&
& \vtop to 0pt{\hbox{\vbox to 7pt{}}\hbox{
$\lambda_{0c}v^2 $ }\vss}\hfil
&
&\cr
height2pt&\omit&&\omit&&\omit&&\omit&\cr
height2pt&\omit&&\omit&&\omit&&\omit&\cr
&  &&  &&  && &\cr
height2pt&\omit&&\omit&&\omit&&\omit&\cr
\omit&\multispan7\hrulefill&\cr
height2pt&\omit&&\omit&&\omit&&\omit&\cr
& \vtop to 0pt{\hbox{\vbox to 7pt{}}\hbox{
$\phi_W\sim\sum_i[t_i(t_i+1)-t^2_{3i}]v_i\phi_i$ }\vss}\hfil
&
& \vtop to 0pt{\hbox{\vbox to 7pt{}}\hbox{
${1\over 4}\lambda_{cc}|G^+|^4$ }\vss}\hfil
&& \vtop to 0pt{\hbox{\vbox to 7pt{}}\hbox{
 $\sum_iv_i^2$ }\vss}\hfil
&
& \vtop to 0pt{\hbox{\vbox to 7pt{}}\hbox{
${1\over 2}\lambda_{cc}v^2 $ }\vss}\hfil
&
&\cr
height2pt&\omit&&\omit&&\omit&&\omit&\cr
height2pt&\omit&&\omit&&\omit&&\omit&\cr
&  &&    && && &\cr
height2pt&\omit&&\omit&&\omit&&\omit&\cr
\omit&\multispan7\hrulefill&\cr
height2pt&\omit&&\omit&&\omit&&\omit&\cr
& \vtop to 0pt{\hbox{\vbox to 7pt{}}\hbox{
${1\over v}\sum_iv_i\phi_i$ }\vss}\hfil
&& \vtop to 0pt{\hbox{\vbox to 7pt{}}\hbox{
${1\over 8}\lambda\phi^4$ }\vss}\hfil
&
& \vtop to 0pt{\hbox{\vbox to 7pt{}}\hbox{
 $\sum_iv_i^2$ }\vss}\hfil
&
& \vtop to 0pt{\hbox{\vbox to 7pt{}}\hbox{
$\lambda v^2 $ }\vss}\hfil&&\cr
height2pt&\omit&&\omit&&\omit&&\omit&\cr
height2pt&\omit&&\omit&&\omit&&\omit&\cr
&  &&  &&  &&  &\cr
height2pt&\omit&&\omit&&\omit&&\omit&\cr
\omit&\multispan7\hrulefill&\cr
height2pt&\omit&&\omit&&\omit&&\omit&\cr
& \vtop to 0pt{\hbox{\vbox to 7pt{}}\hbox{
${1\over v_o}\sum_{i,odd}v_i\phi_i$ }\vss}\hfil
&& \vtop to 0pt{\hbox{\vbox to 7pt{}}\hbox{
${1\over 8}\lambda_o\varphi^4$ }\vss}\hfil
&
& \vtop to 0pt{\hbox{\vbox to 7pt{}}\hbox{
 $\sum_{i,odd}v_i^2$ }\vss}\hfil
&
& \vtop to 0pt{\hbox{\vbox to 7pt{}}\hbox{
$\lambda_ov_o^2 $ }\vss}\hfil&&\cr
height2pt&\omit&&\omit&&\omit&&\omit&\cr
height2pt&\omit&&\omit&&\omit&&\omit&\cr
&  &&  &&  &&  &\cr
height2pt&\omit&&\omit&&\omit&&\omit&\cr
\omit&\multispan7\hrulefill&\cr
}}
$$
\vskip -.15 cm \nobreak
{\baselineskip 10 pt\narrower\smallskip\noindent\ninerm
{\ninebf Table 1:} Summary of mass bounds.
\smallskip}

\vskip .35  cm

In the following we will always assume the equality $\varphi_a=\phi_a$.
The second column gives the relevant quartic couplings by showing the
corresponding piece of the scalar potential. $G^0, G^{\pm}$ are the 
Goldstone bosons. The third column gives the values of the vev's $v_i$
and the fourth lists the scalar form of the mass bounds.

The vector form of the bounds contains extra useful information not contained
in the scalar form. As we have shown, in a particular model one can obtain 
several different mass bounds for the lightest Higgs scalar and by choosing 
the stronger among those one gets the best limit. Here we will show how
the vector form can be used to extract further information on the scalar
spectrum by the interplay between different bounds.

Consider the inequality
$$
\sum_{A}
\langle\phi|A\rangle M_A^2\langle A|\phi\rangle\leq \lambda v^2.
$$
If some state $|A\rangle$ is much heavier than the breaking scale, 
$M_A^2\gg \lambda v^2$, its overlap with $|\phi\rangle$ has to be 
correspondingly small: 
$$
|\langle \phi|A\rangle|^2\leq\lambda v^2/ M_A^2\rightarrow 0.
$$

This can be the case if all mass parameters in the potential are made
very heavy ($\sim M$) while the pattern of symmetry breaking is held fixed 
($v_i$
constant. The required fine-tuning is bigger the heavier the mass scale M
is made). We will call such situation the decoupling limit. In such limit
one expects that all scalars will have masses of the order of the
heavy scale $M\gg v$ with the exception of the state constrained by the  
mass limit $\lambda v^2$ (and possibly others whose mass is protected
by some symmetry). Then, if all scalars but one are very heavy, the light
state $|1\rangle$ will satisfy
$$
|\langle \phi|1\rangle|^2=1-{\cal O}(v^2/M_A^2)\rightarrow 1. $$
This means that the state $|\phi\rangle$ appearing in the vector-form of 
the bound is precisely the state that will remain light in the decoupling 
limit.

\noindent{\bf Exercise.} Given that $|1\rangle\rightarrow|\phi\rangle$ in
the decoupling limit, can you prove that $M_1^2\rightarrow \lambda 
v^2=\langle \phi|M^2|\phi\rangle$?

Suppose now that your model has two `linearly-independent' bounds
$\lambda_1v_1^2=\langle\phi_1|M^2|\phi_1\rangle$
and  $\lambda_2v_2^2=\langle\phi_2|M^2|\phi_2\rangle$ with 
$|\phi_1\rangle\neq\kappa|\phi_2\rangle$. What would be the lightest
state in the decoupling limit, $|\phi_1\rangle$ or $|\phi_2\rangle$?
Of course the way out of this paradox is that the pure decoupling limit
cannot be realized: at least two states will remain light:

\noindent{\bf Exercise.} In the situation described above, prove that
the mass of the second-to-lightest state $M_2$ satisfies the
inequality
$$
M_2^2-M_1^2\leq{1\over\sin^2\alpha}\left\{\sqrt{
\lambda_1v_1^2-M_1^2}+\sqrt{\lambda_2v_2^2-M_1^2}\right\}^2,
$$
where $\alpha$ is the angle between $|\phi_1\rangle$ and $|\phi_2\rangle$.

The existence of a bound on the mass of the second to lightest scalar
in such a situation can be easily understood in terms of the following
geometrical construction. Consider the mass ellipsoid in the 
(multidimensional) scalar field space. Its axis lie along the directions of
the eigenvalues with length proportional to the inverse of the 
corresponding squared mass eigenvalues. One mass bound of the form we have
considered is given in this picture by some vector with length $(\lambda 
v^2)^{-1}$ and direction determined by $|\phi\rangle$. The fact that the 
bound is satisfied implies that the ellipsoid can at most touch the tip
of this vector but cannot intersect it: $\langle \phi |M^2|\phi\rangle\leq
\lambda v^2$. This is represented in the 2D example of \fig\ellips\ a. As is 
clear form the picture
there is necessarily an eigenvalue with mass $M_h^2\leq\lambda v^2$. 
By choosing all mass parameters heavy while keeping $v$ fixed we can flatten
the ellipsoid (i.e. make the mass eigenvalues heavy) except in the 
vector direction of the mass bound, where the flattening is obstructed by
the presence of the vector bound, that the ellipsoid cannot intersect. This
is shown in fig.~\ellips\ b. From it, is clear that when all mass 
eigenvalues but one
are heavy, the light state is given by $|\phi\rangle$. Finally, in the 
presence of two linearly independent mass bounds, the ellipsoid can be 
flattened at most to an ellipse (see fig.~\ellips\ c). The two light 
states, undetermined in general, will be linear combinations of 
$|\phi_1\rangle$ and $|\phi_2\rangle$. 
Obviously, this can be generalized to any number of independent 
mass bounds
and one will expect always a number of light states in correspondence with
the dimensionality of the vector space $\{|\phi_a\rangle\}$.

\vskip .3 cm
\centerline{\epsfxsize 7. truecm \epsfbox{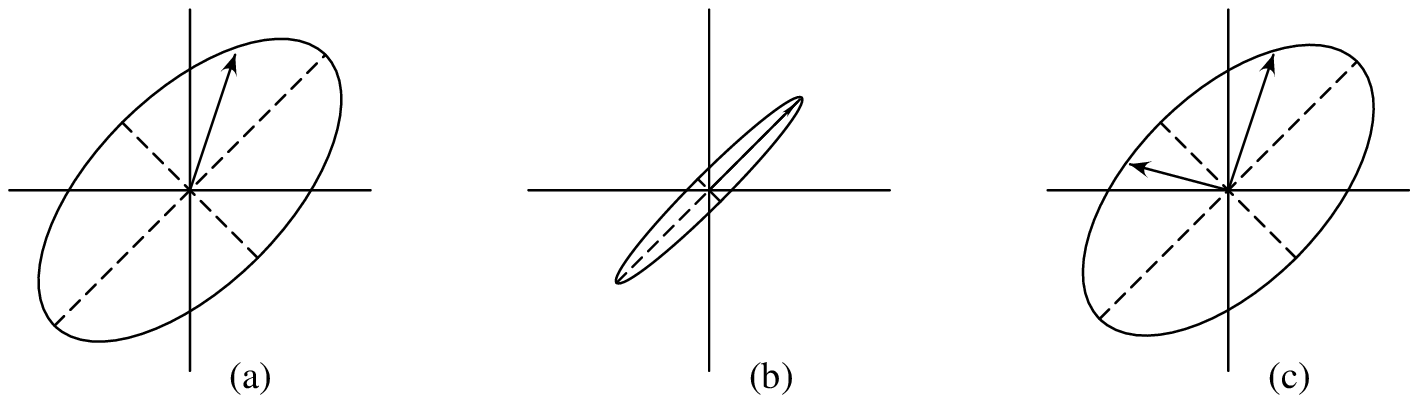} }
\nobreak
\vskip .1 cm
{\baselineskip 10 pt\narrower\smallskip\noindent\ninerm\nobreak
{ Fig.~\ellips :} Geometrical representation of the working of 
mass bounds. \smallskip}

\vskip .4 cm

\noindent{\bf Exercise.} Suppose that there is some `triplet impurity' in 
$SU(2)_L\times U(1)_Y$ breaking, i.e. some $SU(2)$ triplet takes a vev 
$x$ and contributes to the weak vector boson masses in addition to the usual 
doublets. The vev $x$ is bounded to be small by the $\Delta\rho$ constraint.
In the decoupling limit, what triplet admixture do you expect in the light
Higgs boson? Of which order would be the second to lightest Higgs mass? 
\vskip .2cm

\noindent
{\it \Spontex.2 Production Cross Section of the light Higgs boson}
\vskip .1 cm

Concerning electroweak symmetry breaking, besides the knowledge about  
limits on light Higgs masses one would be interested in determining
whether such states can be detected experimentally at all. It is here 
that the information on the composition of the light Higgs becomes
crucial because it determines directly the production and decay cross
sections. 
Here we will limit ourselves as an example, to the Higgs-strahlung 
production mechanism of a light Higgs boson in a $e^+e^-$ collider:
$e^+e^-\rightarrow Z^*\rightarrow Z h_1$. 
The cross section for this process is proportional to the corresponding
cross section for a Standard Model Higgs boson of the same mass. The 
proportionality coefficient is fixed by the gauge properties of $h_1$.
More precisely, the coefficient is given by the overlap between $h_1$
and the $|\phi_Z\rangle$ state listed in table 1:
$$
|\langle \phi_Z|h_1\rangle|^2={\sigma(e^+e^-\rightarrow Zh_1)\over
\sigma(e^+e^-\rightarrow Zh_{SM})}.
$$
Suppose that the model contains a scalar Higgs which is a gauge singlet. 
If the light state $h_1$ has a large overlapping with the singlet the
production cross section for $h_1$ will be reduced. In this situation
even if $h_1$ is forced to be below some mass bound it may be very 
difficult to produce and detect in accelerators.
However, the important point is that from the vector form of the first bound 
listed in table~1 some state with a non-vanishing overlap with $\phi_Z$ 
must remain 
light. In other words, if the lightest Higgs turns out to be orthogonal to
$\phi_Z$ one can still use the information from $\langle 
\phi_Z|M^2|\phi_Z\rangle$ to put a bound on some other Higgs. Technically 
this is realized in the following way: consider the quantities 
$\langle \phi_Z|M^2-m_N^2|\phi_Z\rangle$, where $N$ numbers the 
(scalar) Higgs mass eigenvalues, $m_1\leq m_2...$.
For $N=1$ one obtains
$$\langle \phi_Z|M^2|\phi_Z\rangle=\lambda v^2\leq 0,
$$
which is the original bound. For $N=2$ the following inequality results
$$
m_2^2\leq{\lambda v^2 - m_1^2 |\langle \phi_Z|h_1\rangle |^2\over
1-|\langle \phi_Z|h_1\rangle |^2}.
$$
From this relation we see that if $h_1$ becomes singlet dominated 
$|\langle \phi_Z|h_1\rangle |\rightarrow 0$ and $m_2^2\leq \lambda v^2$,
i.e. the second to lightest eigenvalue satisfies the original bound.
If on the other hand $h_1\rightarrow \phi_Z$ then no bound on $m_2$ can
be set.
In general, for the $N^{th}$ eigenvalue one finds
$$
m_N^2\leq{\lambda v^2 - m_1^2 S_N^2\over
1-S_N^2},
$$
with 
$$
S_N^2=\sum_{p=1}^{N-1}|\langle \phi_Z|h_p\rangle |^2.
$$
When $S_N$ is small, for  the first $N-1$ light Higgses having reduced
couplings to the Z, the bound on $m_N^2$ is then stronger.
This effect can ensure sufficient production of some light scalar provided 
there are not too many singlets.

\section{The lightest Higgs boson in the MSSM}
\tagsection\MSSM

\noindent
{\it \MSSM.1 The MSSM Higgs sector at tree level}
\vskip .1 cm

Supersymmetry requires that the minimal supersymmetric extension of the 
Standard Model, MSSM, contains two Higgs doublets (of opposite 
hypercharge) to give masses to all quarks and charged leptons. 
The most general tree level potential for the $H_1,H_2$ Higgs doublets, 
gauge invariant and renormalizable is then:
$$
\eqalign{
V&=m_1^2|H_1|^2+m_2^2|H_2|^2+[m_{12}^2H_1\cdot H_2+h.c.]\cr
&+{1\over 2}\lambda_1|H_1|^4+{1\over 2}\lambda_2|H_2|^4+
\lambda_3 |H_1|^2|H_2|^2+\lambda_4|H_1\cdot H_2|^2\cr
&+\left[{1\over 2}\lambda_5(H_1\cdot H_2)^2+
\lambda_6 |H_1|^2(H_1\cdot H_2)+\lambda_7 |H_2|^2(H_1\cdot H_2)
+h.c.\right],}
\eqn\vmssm
$$
with 
$$
H_1=\left(\eqalign{&H_1^0\cr &H_1^-}\right),\;\;\;
H_2=\left(\eqalign{&H_2^+\cr &H_2^0}\right).
$$
The quartic couplings in this potential are constrained by supersymmetry 
to be
$$
\lambda_1=\lambda_2={1\over 4}(g^2+g'^2),\;\;\;
\lambda_3={1\over 4}(g^2-g'^2),\;\;\;
\lambda_4=-{1\over 2}g^2,\;\;\; \lambda_5=\lambda_6=\lambda_7=0,
\eqn\susylamb
$$
i.e. they are given in terms of the gauge coupling constants.
The projection of (\use\vmssm) on the neutral Higgs components
gives the potential
$$
\eqalign{
V(H_1^0,H_2^0)&=m_1^2|H_1^0|^2+m_2^2|H_2^0|^2+[m_{12}^2H_1^0 H_2^0+h.c.]\cr
&+{1\over 8}(g^2+g'^2)(|H_1|^2-|H_2^0|^2)^2.}
\eqn\vho
$$
The minimum of this potential determines the vevs
$\langle H_1^0\rangle=v_1/\sqrt{2}$ and $\langle H_2^0\rangle=v_2/\sqrt{2}$ 
with $v_1^2+v_2^2=(246\ GeV)^2$ fixed by the gauge boson masses. 
The ratio $\tan\beta=v_2/v_1$ is a free parameter. The Higgs spectrum
in the broken minimum just described consists of two ($CP$ even) scalars
$h^0, H^0$, one ($CP$ odd) pseudoscalar $A^0$ and two charged Higgses
$H^\pm$. Two of the three mass parameters in (\use\vho) can be traded by
$v_1$ and $v_2$, so that, at tree level, the properties of the Higgs sector
(masses, mixing angles and couplings) are determined by one mass parameter
(usually taken to be the mass of the pseudoscalar, $m_A$) and $\tan\beta$.

The discussion of Section~\use\Spontex\ should have made clear that 
relations (\use\susylamb) have direct and important consequences
for the Higgs spectrum. The tree level masses for Higgs bosons are:
$$
\eqalign{
m_{H^{\pm}}^2&=m_W^2+m_A^2,\cr
m_{h,H}^2&={1\over 2}\left[m_A^2+m_Z^2\mp
\sqrt{(m_A^2+m_Z^2)^2-4m_Z^2m_A^2\cos^22\beta}
\right].}
\eqn\masses
$$
The two neutral Higgses are linear combinations of 
$H_{1,2}^{0r}\equiv\sqrt{2}Re H_{1,2}^0$, see \fig\mssm\ :
\vskip .3 cm
\centerline{\epsfxsize 7. truecm \epsfbox{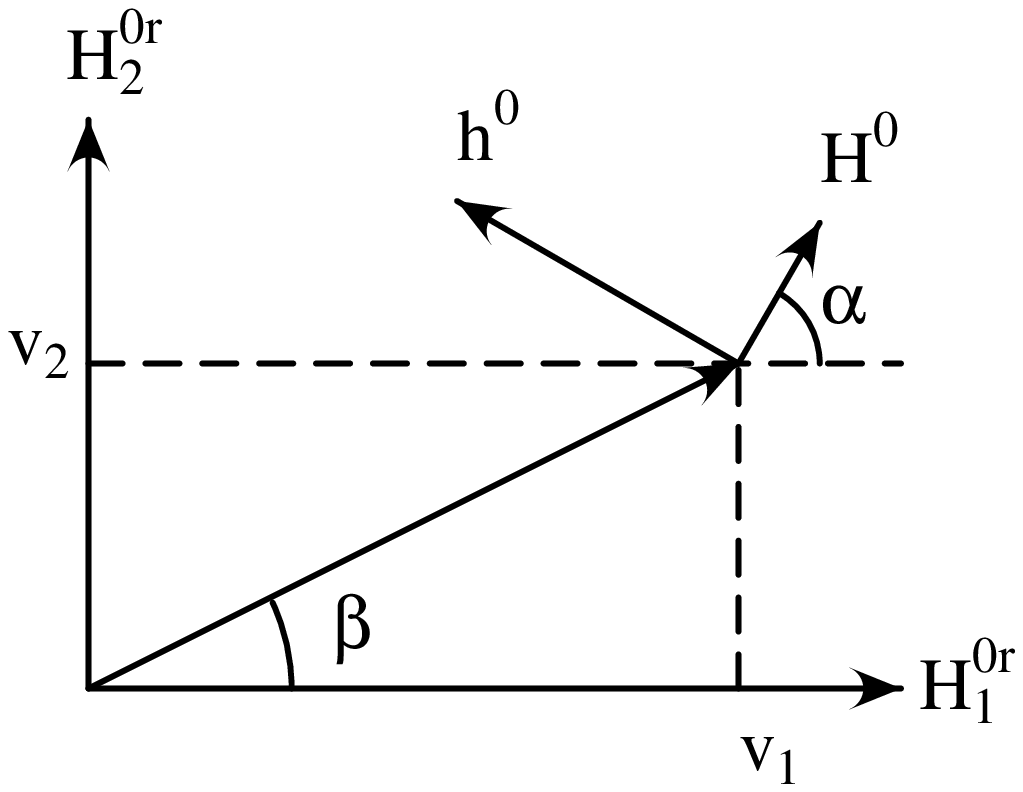} }
\nobreak
\vskip .1 cm
{\baselineskip 10 pt\narrower\smallskip\noindent\ninerm\nobreak
{ Fig.~\mssm :} Electroweak symmetry breaking in the MSSM. 
\smallskip}

\vskip .4 cm
$$
\eqalign{H^0&=(H_1^{0r}-v_1)\cos\alpha+(H_2^{0r}-v_2)\sin\alpha,\cr
h^0&=(H_2^{0r}-v_2)\cos\alpha-(H_1^{0r}-v_1)\sin\alpha,
}
$$
where the mixing angle $\alpha$ is given by
$$
\cos 2\alpha=-\cos 2\beta\,\,{m_A^2-m_Z^2\over m_H^2-m_h^2},\;\;\;
\sin 2\alpha=-\sin 2\beta\,\,{m_H^2+m_h^2\over m_H^2-m_h^2}.
$$
The couplings of $H^0$ and $h^0$ relative to the Standard Model Higgs boson
are then determined by $\alpha$ and $\beta$ as shown in table 2.

\vskip -.2 truecm

$$
\vbox{\offinterlineskip
\halign{ &\vrule# & \strut\quad\hfil#\quad\cr
\omit&\multispan9\hrulefill&\cr
height2pt&\omit&&\omit&&\omit&&\omit&&\omit&\cr
& \vtop to 0pt{\hbox{\vbox to 7pt{}}\hbox{ }\vss}\hfil &
& \vtop to 0pt{\hbox{\vbox to 7pt{}}\hbox{$WW,ZZ$ }\vss}\hfil
&& \vtop to 0pt{\hbox{\vbox to 7pt{}}\hbox{$ZA$ }\vss}\hfil
&& \vtop to 0pt{\hbox{\vbox to 7pt{}}\hbox{$t\overline{t}$ }\vss}\hfil
&& \vtop to 0pt{\hbox{\vbox to 7pt{}}\hbox{$b\overline{b},\tau^+\tau^-$ 
}\vss}\hfil &&\cr
height2pt&\omit&&\omit&&\omit&&\omit&&\omit&\cr
height2pt&\omit&&\omit&&\omit&&\omit&&\omit&\cr
&  && && && && &\cr
\omit&\multispan9\hrulefill&\cr
height2pt&\omit&&\omit&&\omit&&\omit&&\omit&\cr
\omit&\multispan9\hrulefill&\cr
height2pt&\omit&&\omit&&\omit&&\omit&&\omit&\cr
& \vtop to 0pt{\hbox{\vbox to 7pt{}}\hbox{
$h^0$ }\vss}\hfil
&& \vtop to 0pt{\hbox{\vbox to 7pt{}}\hbox{
$\sin(\beta-\alpha)$ }\vss}\hfil
&& \vtop to 0pt{\hbox{\vbox to 7pt{}}\hbox{ $\cos(\beta-\alpha)$ }\vss}\hfil
&& \vtop to 0pt{\hbox{\vbox to 7pt{}}\hbox{
$\cos\alpha/\sin\beta $ }\vss}\hfil
&& \vtop to 0pt{\hbox{\vbox to 7pt{}}\hbox{
$-\sin\alpha/\cos\beta $ }\vss}\hfil
&&\cr
height2pt&\omit&&\omit&&\omit&&\omit&&\omit&\cr
height2pt&\omit&&\omit&&\omit&&\omit&&\omit&\cr
&  && &&   && && &\cr
height2pt&\omit&&\omit&&\omit&&\omit&&\omit&\cr
\omit&\multispan9\hrulefill&\cr
height2pt&\omit&&\omit&&\omit&&\omit&&\omit&\cr
& \vtop to 0pt{\hbox{\vbox to 7pt{}}\hbox{
$H^0$ }\vss}\hfil
&& \vtop to 0pt{\hbox{\vbox to 7pt{}}\hbox{
$\cos(\beta-\alpha)$ }\vss}\hfil
&& \vtop to 0pt{\hbox{\vbox to 7pt{}}\hbox{ $\sin(\beta-\alpha)$ }\vss}\hfil
&& \vtop to 0pt{\hbox{\vbox to 7pt{}}\hbox{
$\sin\alpha/\sin\beta $ }\vss}\hfil
&& \vtop to 0pt{\hbox{\vbox to 7pt{}}\hbox{
$\cos\alpha/\cos\beta $ }\vss}\hfil
&&\cr
height2pt&\omit&&\omit&&\omit&&\omit&&\omit&\cr
height2pt&\omit&&\omit&&\omit&&\omit&&\omit&\cr
& &&  &&  &&  && &\cr
height2pt&\omit&&\omit&&\omit&&\omit&&\omit&\cr
\omit&\multispan9\hrulefill&\cr
}}
$$
\vskip -.15 cm \nobreak
{\baselineskip 10 pt\narrower\smallskip\noindent\ninerm
{Table 2: Neutral MSSM Higgs couplings relative to the 
corresponding Standard Model Higgs coupling:  $g_{SUSY}/g_{SM}$}. \smallskip}

\vskip .35  cm

We can apply some of the general results examined in section 2
to the MSSM case just described. For instance, if the only mass parameter
available in the Higgs sector, $m_A$, is made heavy (compared with the 
electroweak scale as given by $M_Z$), eqs. (\use\masses) show that all
Higgses will acquire masses of order $m_A$ except one neutral Higgs that
remains light, with
$$
m_h^2\leq M_Z^2\cos^22\beta.
$$
The inequality holds for any value of $m_A$ and is saturated when 
$m_A\rightarrow \infty$. This bound explicitly derived here is the 
particularization to the MSSM  of the general bound (\use\susyb) and is 
associated with a vector $|\varphi\rangle$ in $(H^0,h^0)$ space that lies 
along the direction of the breaking, see fig.2. In the decoupling limit, 
$m_A\gg m_Z$, we know that the light state $h^0$ will tend to be aligned 
with the vector $|\varphi\rangle$. That is, $\alpha\rightarrow 
\beta-\pi/2$ or 
$$
h^0\sim (v_1 h_1^0+v_2h_2^0)/v=  h_1^0\cos\beta 
+h_2^0\sin\beta .
\eqn\ho
$$ 
In addition, a look at table 2 shows that the 
couplings of the light state $h^0$ tend to the Standard Model values in 
this decoupling limit. Such a Higgs scalar should be detectable then at LEPII
(when $m_A\sim m_Z$ the coupling $h^0ZZ$ can be suppressed 
closing the
Higgs-strahlung channel but then the complimentary production mechanism
$e^+e^-\rightarrow Z^*\rightarrow h^0A^0$ becomes important).

In summary, a tree level analysis of the MSSM Higgs sector predicts the 
existence of a CP even Higgs scalar with mass below $M_Z$ that should be
detectable at LEPII. This would represent a stringent test for the 
simplest realization of the supersymmetric extension of the Standard 
Model. However, as is well known, the effect of 
radiative corrections modifies this expectation in a dramatic way. 
\vskip .2cm

\noindent
{\it \MSSM.2 Radiatively corrected $m_{h^0}$. Dominant effect}
\vskip .1 cm

After including radiative corrections, the mass of the MSSM lightest Higgs 
boson is no longer determined only by $m_A$ and $\tan\beta$ but will 
depend on the rest of parameters of the theory. In particular, the most 
important corrections come from top-stop loops that shift the squared mass
of $h^0$ by
$$
\delta m_{h^0}^2\sim g^2 {M_t^4 \over M_W^2}\log{M_{\tilde t}^2\over M_t^2}.
\eqn\shift
$$
Here $M_t$ and $M_{\tilde t}$ are the top and stop masses respectively.
Some comments are in order. Note first that the contribution cancels if top 
and stops were degenerate as would correspond to the supersymmetric 
limit. Also note the strong dependence with $M_t$. As an example, if 
$M_t=170\ GeV$, $M_{\tilde t}=1\ TeV$, the maximum tree level value $m_{h^0}=
M_Z$ would be shifted by (\use\shift) by as much as $30\ GeV$ (the 
increase can be even more dramatic when the tree level mass is smaller).

The radiatively corrected Higgs sector of the MSSM has been intensively
studied during the last years using three main tools: diagrammatic 
calculations, effective potential techniques and renormalization group 
methods.
It is instructive to obtain (\use\shift) using these three approaches.

\noindent{\bf Exercise.} Calculate the dominant correction (\use\shift) 
diagrammatically. Assume 
for simplicity $m_A\gg m_{h^0}$, i.e. work only with $h^0$ as defined in 
(\use\ho). Consider the diagrams of \fig\dia\ using the rules $h^0 
\overline{t}t$: $H_t/\sqrt{2}$; $h^0h^0{\tilde t}^*_{L,R}{\tilde t}_{L,R}$:
$H_t^2$, where $H_t\equiv h_t\sin\beta$. Also, $M_t^2=H_t^2v^2/2$,
$M_{{\tilde t}_L}^2=m_Q^2+M_t^2+{\cal O}(g^2v^2)$,
$M_{{\tilde t}_R}^2=m_U^2+M_t^2+{\cal O}(g^2v^2)$. [Left-right mixing in the
stop sector can be neglected to compute the dominant effect (\use\shift)].
\vskip .3 cm
\centerline{\epsfxsize 10. truecm \epsfbox{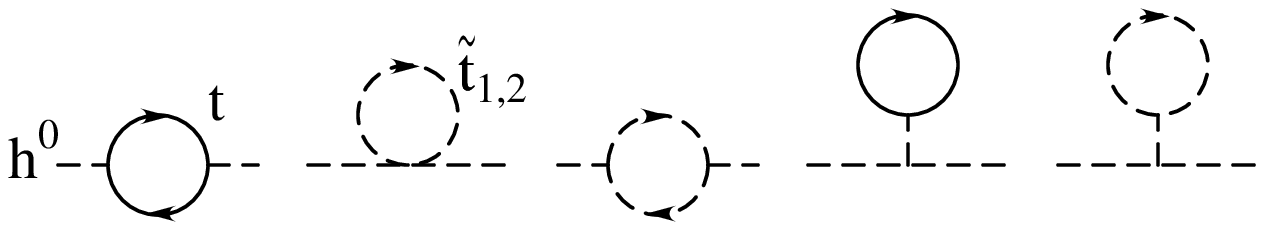} }
\nobreak
\vskip .1 cm
{\baselineskip 10 pt\narrower\smallskip\noindent\ninerm\nobreak
{ Fig.~\dia :} Top and stop one-loop diagrams contributing to $m_{h^0}^2$. 
\smallskip}

\vskip .4 cm

\noindent{\bf Exercise.} The one-loop MSSM effective potential when 
$m_A\gg m_Z$ is given by $$
\eqalign{
V(h)&=-{1\over 2} m^2h^2+{1\over 8}\lambda h^4\cr
&+{6\over 64 \pi^2}\left[M_{{\tilde t}_1}^4(\log{M_{{\tilde t}_1}^2\over 
\mu^2}-{3\over 2})+M_{{\tilde t}_2}^4(\log{M_{{\tilde t}_2}^2\over 
\mu^2}-{3\over 2})-2M_t^4(\log{M_t^2\over 
\mu^2}-{3\over 2}) \right],
}
$$
with $\lambda=[(g^2+g'^2)/4]\cos^22\beta$ and $M_{{\tilde t}_{1,2}}^2=
M_{{\tilde t}_{L,R}}^2$ if left-right mixing is neglected.
The second derivative of the potential in the minimum $\langle\phi\rangle=v$
will give the one-loop corrected Higgs mass.

The third method, perhaps the most elegant, is based on an effective field 
theory approach: when the SUSY spectrum is around some scale $M_S\gg M_Z$
(in particular $M_{{\tilde t}}\sim m_A\sim M_S$) the effective theory below
$M_S$ is just the Standard Model. In particular remember that $m_A\gg M_Z$
implies that the light Higgs boson [given by (\use\ho)] has Standard 
Model-like properties.
The quartic coupling of $h^0$ is determined by supersymmetry at the scale 
$M_S$:
$$
\lambda (M_S)={1\over 4}[g^2(M_S)+g'^2(M_S)]\cos^22\beta,
\eqn\boundary
$$ 
and its value at the electroweak scale can be computed integrating the 
renormalization group equations in the effective theory between $M_S$ and 
the low energy scale (e.g. $M_t$) with (\use\boundary) as a boundary 
condition.
The mass of $h^0$ can then be obtained from $\lambda$ at the weak 
scale by using the Standard Model relation $m_h^2=\lambda v^2$.
It is then straightforward to reinterpret the result (\use\shift):
the logarithmic dependence arises from the running of $\lambda$ from
$M_{\tilde t}\sim M_S$ to $M_t$ and the $M_t^4$ dependence comes from
the dominant piece of the Standard Model beta function for $\lambda$:
$$
\beta^{SM}_{\lambda}\sim -{12\over 16\pi^2}h_t^4.
$$
Note that this renormalization group method is particularly well suited 
to study the upper bound on $m_{h^0}^2$ which is obtained in the 
large $m_A$ limit.

A few comments on the structure of the radiative corrections to $m_{h^0}^2$
are in order. The fact that one-loop radiative corrections to this mass 
can be very sizeable (and even larger than the tree level mass if $\tan\beta$
is small) does not mean that perturbation theory is not reliable. A large
ratio of one-loop corrections to tree-level contributions arises because
the tree-level result does not depend on the large top Yukawa coupling while
one-loop corrections do. Furthermore the one-loop result is enhanced by 
a logarithm of a possibly heavy mass ($M_S$) to a light mass ($M_t$). The 
loop contributions in the adimensional ratio $\Delta m_{h^0}^2/M_t^2$ are
basically of the form
$$
\sum_n\sum_{k=0}^n (\alpha\log)^k\alpha^{n-k},
$$
where $\alpha=h_t^2/4\pi^2$ and $\log\sim\log(M_S^2/M_t^2)$. The 
terms for a given $n$ come from the n-loop corrections, with the
index $k$ corresponding to hierarchically organized contributions 
(provided the $\log$ is sizeable): leading-log terms ($k=n$), 
next-to-leading ($k=n-1$) and so on. For the 
perturbative expansion to be reliable we should require $\alpha\log<1$
which is satisfied for the current top mass values.
The use of renormalization group techniques permits to reorganize the
loop expansion resumming to all loops the numerically most important
corrections (leading, next-to-leading, etc). In the next subsection we
discuss how to implement a computation of the loop corrected mass $m_{h^0}$
that will include up to next-to-leading radiative corrections. Use
will be made of the three methods just sketched above for the dominant
correction.
\vskip .2cm

\noindent
{\it \MSSM.3 Radiatively corrected $m_{h^0}$. Next-to-leading log 
computation} 
\vskip .1 cm

The ingredients for a next-to-leading log computation of the radiatively 
corrected $m_h^0$ will be discussed in this 
subsection. We assume that the supersymmetric spectrum can be described
by a common mass $M_S$ well above the electroweak scale. In particular
$m_A\sim M_{\tilde t}\sim M_S$ and below $M_S$ the effective theory is
the Standard Model. As we saw, the quartic Higgs coupling at the weak scale 
will determine the light Higgs mass, while at the supersymmetric scale its
value is fixed by supersymmetric parameters. 
First, we can see that the integration of the coupling $\lambda$ from $M_S$ 
down to the electroweak scale indeed resums some series of corrections to 
all loops. For example, from 
$d\lambda/dt=\beta_\lambda$, where $t=\log\mu$, we get
$$
\eqalign{
\lambda (M_S)&-\lambda (M_t)=\int_{\mu=M_t}^{M_S}\beta_\lambda(t)dt\cr
&=\int_{\mu=M_t}^{M_S}\left[
\beta_\lambda(t_0)+(t-t_0){d\beta_\lambda\over dt}(t_0)+...+
{1\over n!}(t-t_0)^n{d^n\beta_\lambda\over dt^n}+...
\right]dt\cr
&=\beta_\lambda(t_0)\log{M_S\over M_t}+{1\over 2}{d\beta_\lambda\over 
dt}(t_0)\left[\log{M_S\over M_t}\right]^2+...+{\cal O} 
\left[\log{M_S\over M_t}\right]^n+...}. 
$$
Inserting in the above expression a loop expansion for the beta function
one recovers the general estructure of the radiative corrections discussed
at the end of the previous subsection. The one-loop approximation for 
the beta functions corresponds to the (all loop) leading log contributions.
Using two loop beta functions would resum also the next-to-leading logs
and so on.

The fact that information on higher loop corrections can be obtained with 
the knowledge of just one-loop beta functions is due to the magic of the 
Renormalization Group. Let us have a closer look to it from the point of
view of the effective potential $V(\varphi)$. The starting point is the 
observation that
a change in the renormalization scale $\mu\rightarrow\mu+d\mu$ doesn't 
change the physics.
The invariance of the potential under such change
$
V(\varphi(\mu),\lambda_i(\mu);\mu)=
V(\varphi(\mu+d\mu),\lambda_i(\mu+d\mu);\mu+d\mu),
$
can be expressed in differential form as (sum over 
repeated indices implied)
$$
\left[
\beta_i{\partial\over\partial\lambda_i}-
\gamma\varphi{\partial\over\partial\varphi}+
\mu{\partial\over\partial\mu}
\right]V=0.
\eqn\rgv
$$
Here $\lambda_i$ stands for a generic coupling with corresponding beta
function $\beta_i$, $\varphi$ is the Higgs field, with anomalous dimension
$\gamma$ ($\gamma\varphi=-d\varphi/dt$). Note that beyond tree level the 
effective potential also depends explicitly on $\mu$ through logarithms.
Then (\use\rgv) connects contributions to the potential of different 
orders in $\hbar$. For example, at order $\hbar^1$ one has 
$$
\beta_i^{(1)}{\partial V_0\over\partial\lambda_i}-
\gamma^{(1)}\varphi{\partial V_0\over\partial\varphi}+
\mu{\partial V_1\over\partial\mu}=0,
\eqn\rgvu
$$
where $V_0$ and $V_1$ are the tree-level and one loop potentials respectively
and the index $(1)$ in the $\beta$ and $\gamma$ functions indicate one-loop
approximations. Eq. (\use\rgvu) tells that knowledge of $V_0$, and 
one-loop rg functions allows the computation of the leading-log 
one-loop contribution in $V_1$ (which goes like $\log\mu$).

\noindent{\bf Exercise.} The reciprocal is also true. From the Standard Model
effective potential
$$
V=-{1\over 2}m^2\varphi^2+{1\over 8}\lambda \varphi^4+\sum_i{n_i\over 
64\pi^2}M_i^4(\varphi)\left[
\log{M_i^2(\varphi)\over \mu^2}-C_i
\right] ,
$$ 
with $n_t=-12$, $M_t^2=h_t^2\varphi^2/2$; $n_W=6$, $M_W^2=g^2\varphi^2/4$;
$n_Z=3$, $M_Z^2=(g^2+g'^2)\varphi^2/4$; $n_h=1$, $M_h^2=3\lambda\varphi^2/2-
m^2$; $n_\chi=3$, $M_\chi^2= \lambda\varphi^2/2-m^2$ (Goldstone bosons),
obtain $\beta_{m^2}^{(1)}$ and $\beta_\lambda^{(1)}$ knowing that
$\gamma^{(1)}=3[h_t^2-(1/4)g'^2-(3/4)g^2].$

Defining the operators 
$$
{\cal D}^{(n)}\equiv \beta_i^{(n)}{\partial\over\partial\lambda_i}-
\gamma^{(n)}\varphi{\partial\over\partial\varphi},
$$
we can write the $\hbar^2$ expression of (\use\rgv) as
$$
{\cal D}^{(2)}V_0+{\cal D}^{(1)}V_1+\mu{\partial V_2\over \partial\mu}=0,
$$
which would imply that knowledge of rg functions to two-loop order permits
to obtain the leading and next-to-leading two-loop contributions in $V_2$
(provided $V_0$ and $V_1$ are also known). 

The procedure can be extended to order $\hbar^n$:
$$
{\cal D}^{(n)}V_0+{\cal D}^{(n-1)}V_1+...+{\cal D}^{(1)}V_{n-1}+\mu{\partial 
V_n\over \partial\mu}=0, 
$$
and recursion relations can be written for the nth-loop leading  
$V_n^{LL}$ and next-to-leading log $V_n^{NTLL}$ contributions:
$$
\eqalign{
{\cal D}^{(1)}V^{LL}_{n-1}+\mu{\partial 
V_n^{LL}\over \partial\mu}&=0,\cr
{\cal D}^{(2)}V^{LL}_{n-2}+{\cal D}^{(1)}V^{NTLL}_{n-1}+\mu{\partial 
V_n^{NTLL}\over \partial\mu}&=0.}
$$
From them we see that $V_n^{LL}$ can be obtained from $V_0$ and ${\cal 
D}^{(1)}$ while to obtain $V_n^{NTLL}$ one needs in addition $V_1$ and
${\cal D}^{(2)}$.
The general statement is that the $L^{th}$ loop potential, with parameters
running with $L+1$ rg functions resums contributions up to 
$L^{th}$-to-leading order.
In our particular case we shall use the one-loop effective potential with 
parameters running at two-loops to resum leading and next-to-leading 
corrections. This should be compared with the approximmation in which
the tree level potential with parameters running at one-loop is used, which
would resum only the leading logs.

An approximation for the effective potential truncated at some loop order,
like the one we are using, will have some residual scale dependence. 
Understandably, this dependence will be ameliorated if we use a one-loop 
expression for $V$ with two-loop running parameters as compared with the 
tree-level (one-loop rg improved) potential approximation. The goal is to
compute reliably the second derivative of the potential (to be related 
with the Higgs mass) in the electroweak minimum, so that one would like 
to keep control on the scale dependence in that region of the minimum.

One way of doing this is the following. If the potential were known exactly
it would be exactly scale independent. In such a case it is simple to 
show that the vev $\langle\varphi (t)\rangle$ should run with the scale 
in the same way as the field $\varphi(t)$ does
$$
\varphi(t)=\varphi_c\exp^{-\int_0^t\gamma(t')dt'}\equiv\varphi_c\xi(t),
$$
where $\varphi_c$ is the classical field. Then, the ratio 
$\langle\varphi (t)\rangle/\xi(t)$ gives a measure of the scale independence
of the potential used to extract $\langle\varphi (t)\rangle$: the ratio 
should be flat in the region where the potential is more scale independent.
This fact can be used to determine numerically some scale $t^*$ where indeed
the above ratio turns out to be flat. This is shown in \fig\scalei\ , where
the ratio $\langle\varphi (t)\rangle/\xi(t)$ is plotted as a function of the
scale for two different approximations for the effective potential: the 
dashed line corresponds to the tree-level (one-loop rg improved) effective
potential and the solid line to the one-loop (two-loop rg improved) case.
The figure shows clearly the improvement in scale independence if the
second approximation is used. The scale where the corresponding curve is 
flatter determines the scale $t^*$: not surprisingly it is of the order of
the mass scale involved in the problem i.e. the top mass.

As a reassuring cross check one can also compare the real running of 
second derivatives (at the minimum) of the potential used with the running
if the potential were scale invariant. Again it is simple to prove that
$\partial^n V/\partial\varphi(t)^n$ runs like $\xi^{-n}(t)$ in the latter
case. Then the ratio of $m_{h,der}^2\equiv\partial^2V/\partial\varphi(t)^2$
at $\langle\varphi(t)\rangle$ to $m_h^2(t)\equiv 
m_h^2(t^*)\xi^2(t^*)/\xi^2(t)$ measures the scale independence of the
potential approximation. It turns out that this ratio becomes flatter at 
approximately the same scale $t^*$ previously found which is then used as 
the best and most reliable scale choice for numerical computations.

\vskip .3 cm
\centerline{\epsfxsize 15. truecm \epsfysize 9. truecm \epsfbox{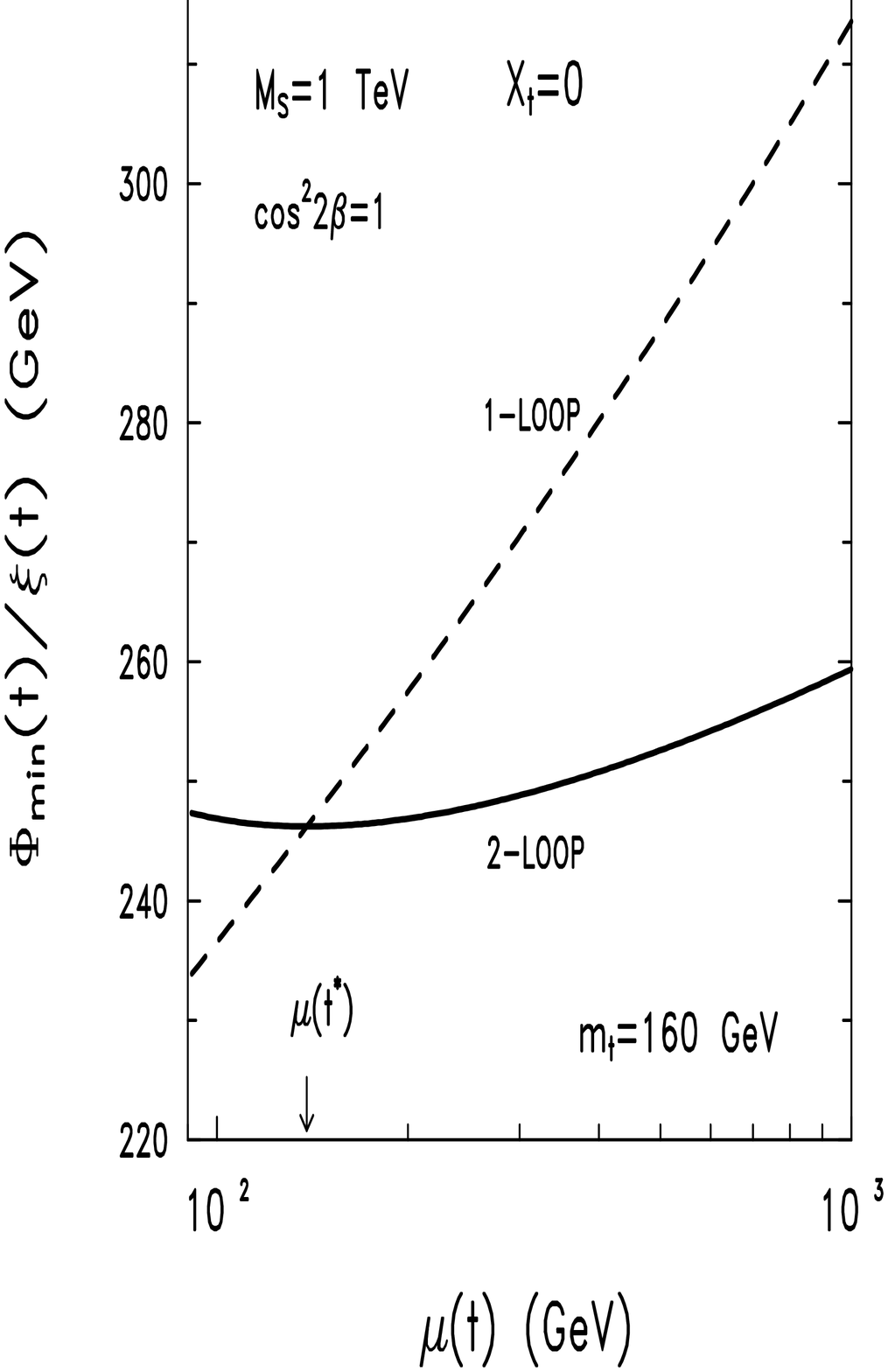} }
\nobreak
{\baselineskip 10 pt\narrower\smallskip\noindent\ninerm\nobreak
{ Fig.~\scalei :} Ratio $\langle\varphi(t)\rangle/\xi(t)$ as a measure of 
scale independence for the effective potential in two different 
approximations as explained in the text. \smallskip}

\vskip .4 cm

Once the scale dependence of the potential is taken care of, one can 
compute the second derivative at the minimum and relate it to the Higgs mass.
This mass will then include logarithmic corrections up to next-to-leading 
order. However one should realize that this is not yet the physical mass, 
i.e. it does not correspond to the pole of the Higgs propagator, 
$\Gamma_R(p^2)=p^2-(m_R^2+\Pi_R(p^2)$, where $p$ is the external momentum,
$m_R$ is the renormalized mass and $\Pi_R$ the one-loop self-energy. 
Remembering that the effective potential generates 1PI diagrams with 
{\it zero} external momentum we get
$$
m_h^2(t)\equiv{\partial^2V\over\partial\varphi^2}
\biggr|_{\langle\varphi\rangle}=-\Gamma_R(p^2=0)=m_R^2+\Pi_R(0).
$$
The pole mass being defined by $\Gamma_R(p^2=M_H^2)=0$, we arrive at
$$
M_H^2=m_h^2(t)+\Pi_R(p^2=M_H^2)-\Pi_R(p^2=0).
\eqn\physh
$$
It can be shown that the scale dependence of $m_h^2(t)$ cancels at 
one-loop with the self-energy difference correction giving rise to a
pole mass scale independent up to higher orders. The diagrammatic 
calculation of the one-loop self-energies adds to the Higgs mass the
one-loop corrections not accesible to rg resummation. 
A similar self-energy correction should be included to relate the running
top mass $m_t(t)=h_t(t)v\xi(t)/\sqrt{2}$ with the top pole mass $M_t$. The 
dominant piece comes from QCD radiative corrections. In $\overline{\rm 
MS}$ it reads
$$
M_t=m_t(\mu=M_t)\left[
1+{4\over 3}{\alpha_s(M_t)\over\pi}\right].
\eqn\physt
$$

The last piece for a consistent computation of $m_{h^0}$ at next-to-leading
order corresponds to the inclusion of one-loop threshold corrections to
the boundary condition (\use\boundary). They arise when integrating out the
heavy supersymmetric particles. The dominant contribution corresponds 
to stops and is represented diagrammatically in \fig\thres\ . It is a 
simple {\bf Exercise} to compute it either by expanding the contribution of 
stops to the MSSM potential
in powers of the background Higgs field or by diagrammatic calculation.
The correction to $\lambda(M_S)$ is proportional to the stop mixing 
$M_{{\tilde t}_L{\tilde t}_R}^2=H_t\varphi X_t$, where 
$H_t=h_t\sin\beta$ and $X_t=(A_t+\mu\cot\beta)$:
$$
\delta\lambda={3 H_t^4\over 8\pi^2}{X_t^2\over M_S^2}\left[
1-{X_t^2\over 12 M_S^2}\right].
\eqn\shiftl
$$

\vskip .3 cm
\centerline{\epsfxsize 10. truecm \epsfbox{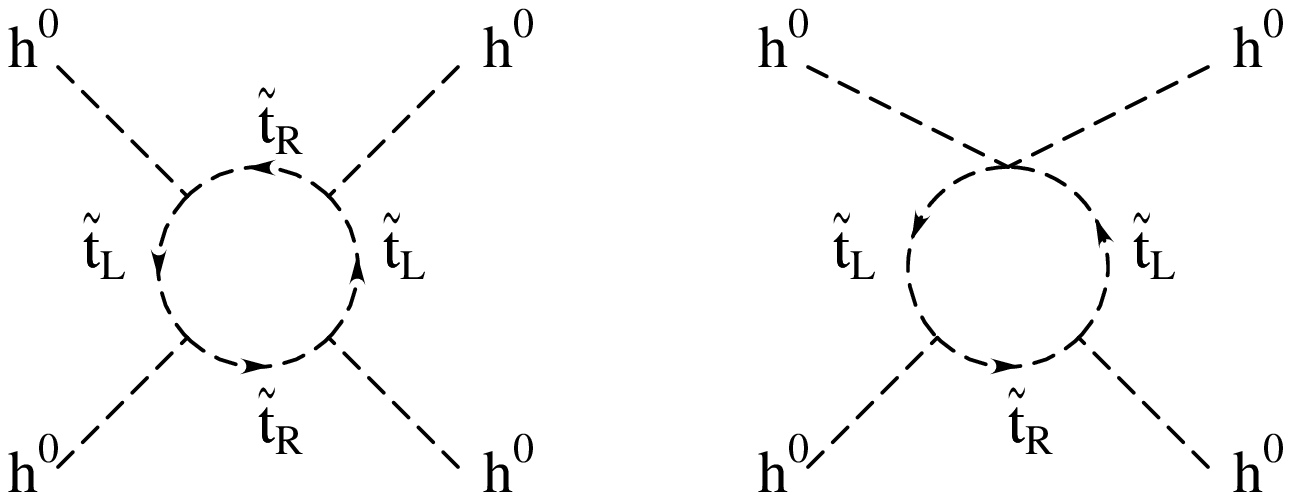} }
\nobreak
\vskip .1 cm
{\baselineskip 10 pt\narrower\smallskip\noindent\ninerm\nobreak
{ Fig.~\thres : Two types of supersymmetric diagrams giving the dominant 
threshold corrections to the quartic Higgs coupling.}  \smallskip}

\vskip .4 cm

The shift in $\lambda$ reaches a maximum value for $X_t^2=6M_S^2$ that
corresponds then to the maximum of the Higgs mass (maximal mixing case).
The case of negligible mixing $X_t\sim 0$ (minimal mixing case)
will in general correspond to the minimum value of the Higgs mass (when 
the rest of parameters is fixed). Note  that the correction 
(\use\shiftl) can be negative if $X_t^2>12 M_S^2$. However in that region of
parameters is easy to run into problems with color or charge breaking 
minima in the full supersymmetric scalar potential.

This completes the list of ingredients for the next-to-leading log 
computation of $m_{h^0}$. An example of the results is plotted in 
\fig\mass\ which gives the (physical) Higgs mass versus the (pole) top 
mass.

\centerline{\epsfxsize 15. truecm \epsfysize 9. truecm \epsfbox{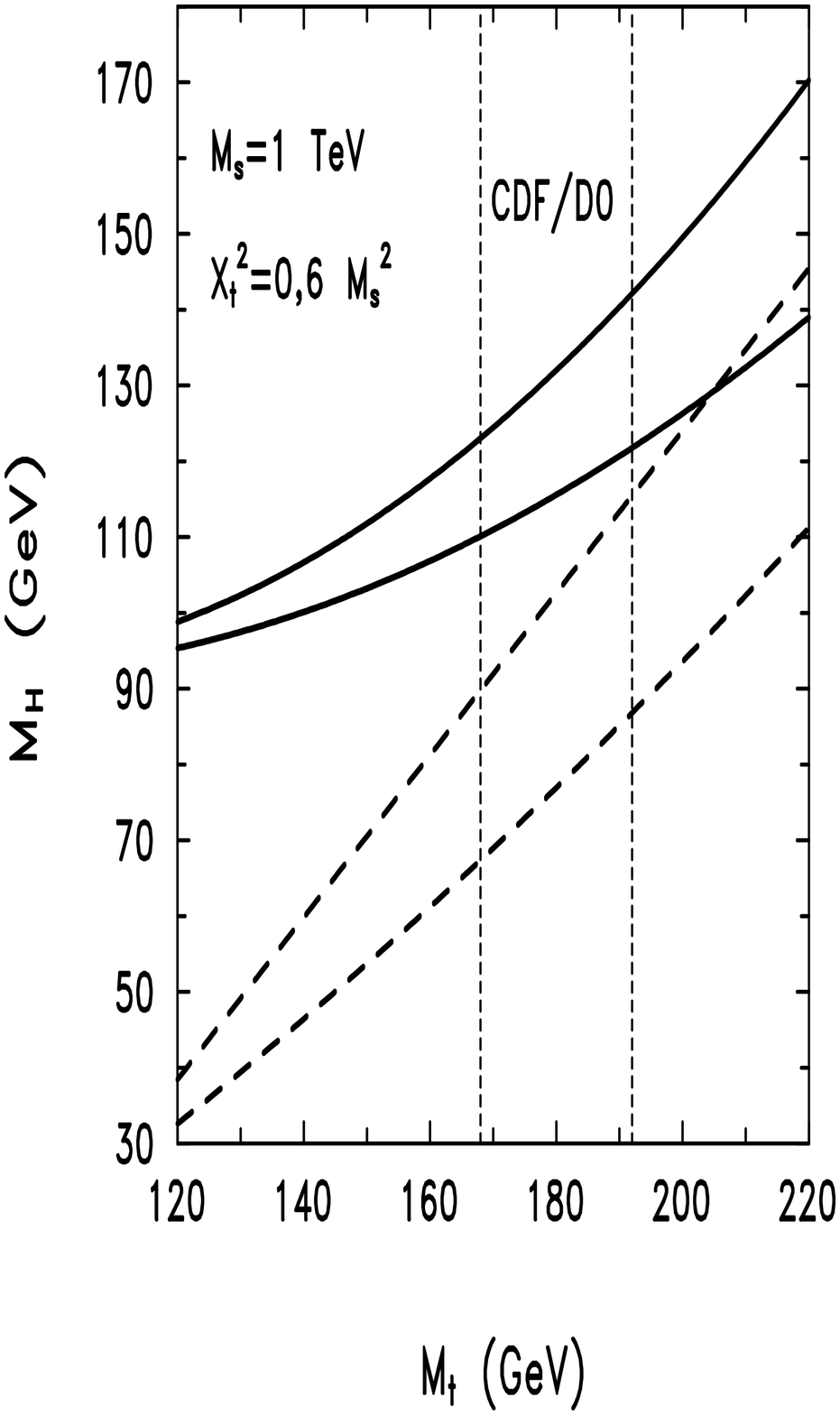} }
\nobreak
\vskip .1 cm
{\baselineskip 10 pt\narrower\smallskip\noindent\ninerm\nobreak
{ Fig.~\mass : Upper limits for the lightest Higgs boson in the MSSM with
the scale of supersymmetry $M_S=1\ TeV$. Solid lines: $\tan\beta\gg 1$,
dashed lines: $\tan\beta=1$. For a given $\tan\beta$ the upper (lower) curve 
corresponds to $X_t^2=6 M_S^2$ $(X_t=0)$.}  
\smallskip}

\vskip .4 cm
\noindent The scale of the supersymmetric particles is set to $1\ 
TeV$ which is roughly the upper limit from naturality arguments. As the 
pseudoscalar mass $m_A$ is then much larger than the electroweak scale,
the masses shown in the figure are actually the upper limits 
for $m_{h^0}$ (for the corresponding value of $M_S$. The masses
do increase logarithmically with $M_S$). Solid lines correspond to the 
case of large $\tan\beta$ while dashed lines have $\tan\beta=1$. In 
each pair, the upper curve is the one for maximal 
mixing $X_t^2=6M_S^2$ (giving then the absolute bound on $m_{h^0}$) and 
the lower for the minimal mixing case $X_t\sim 0$. Short-dashed 
vertical lines give the CDF/D0 range for the top mass. From the figure is 
clear that LEPII can miss the lightest Higgs boson if e.g. $\tan\beta$ 
turns out to be large and stops are heavy with substantial mixing. 

A few comments are in order before closing this section. When there is still
a hierarchy between $M_S$ and $M_t$ but $m_A\sim M_t$ one can repeat the
above procedure with the difference that now, the effective theory below
$M_S$ is a two Higgs doublet extension of the Standard Model. The Higgs 
potential is then that written in (\use\vmssm) with quartic couplings
fixed by supersymmetry at the scale $M_S$. The rg program uses now rg 
functions
for the two Higgs doublet model to evolve the mass matrix for neutral 
scalars,
plus the masses for charged Higgs bosons from
$M_S$ down to the electroweak scale. The mass of the lightest Higgs boson
is smaller in this case.

The fact that a clever choice of the renormalization scale permits to 
obtain reliable results in a leading log calculation can be used to derive
simple analytical formulas for radiatively corrected Higgs masses (and 
couplings). This can be done by iteratively integrating rg equations to 
the required precision.

\section{Standard Model Stability Bound on the Higgs Mass}
\tagsection\Stab

In section 3 we saw that in the MSSM large radiative corrections
to the lightest (CP even) Higgs mass arise in the case of a heavy
supersymmetric scale. By using effective theory methods, the corrections 
were described as a renormalization group effect: the quartic Higgs coupling
at the supersymmetric scale is small (as it is fixed by electroweak gauge 
couplings) but is driven to large values at the electroweak scale by the top 
quark contributions to $\beta_\lambda\sim -12h_t^4/(16\pi^2)$ which dominate
the running of $\lambda$ if the top is heavy. 

This effect of top loop corrections is a purely Standard Model effect (the
running of $\lambda$ below the supersymmetric scale is described by the
Standard Model rg functions) and has interesting consequences already in
the pure Standard Model. The steepness in the running of $\lambda$ implies
that if this quartic coupling is small at the electroweak scale (this means
a light Higgs mass) it can be driven to negative values at a large scale
if the top is sufficiently heavy. A negative value of $\lambda$ signals
the appeareance of an instability in the effective potential at large
values of the field: $V(\varphi)\sim \lambda\varphi^4\rightarrow -\infty$.
If this pathology would appear for values of the field well beyond the
Planck mass we shouldn't worry about it because we know already that the
Standard Model has to be modified at such energy scales. In general, if
the Standard Model is valid up to some large energy scale $\Lambda$ 
(where some new physics will take over) we should be concerned about the 
possibility that the effective potential is destabilized below $\Lambda$.
To avoid this pathology the Higgs mass should be heavy enough
$$
M_H>M_H^c(M_t,\Lambda),
$$
where the critical value $M_H^c$ will be a monotonically increasing function
of the top mass (the heavier the top is, the steeper the descent of $\lambda$
will be) and the scale $\Lambda$ (the larger $\Lambda$ is, the longer will be
the running of $\lambda$).
The interesting point is that, for the current CDF/D0 values of the top mass
the limits $M_H^c$ are around $100\ GeV$, a region with direct 
significance for future Higgs searches. It would then be desirable to compute
this critical masses, dubbed stability bounds, with good precision and this
can be done following a procedure very similar to the one used in the
previous section to compute the upper bound for the lightest MSSM Higgs 
boson mass.

Before that, let us have a closer look at the instability of the potential.
We need to compute reliably the value of the effective potential at large
values of the field to see whether it is below the value at the electroweak
minimum, in which case this minimum would get destabilized and eventually 
decay. As we don't know the exact potential we have to rely in 
perturbative approximations and so, a convenient choice of the 
renormalization scale should be made. To start just consider the choice 
$\mu\sim M_Z$ that has the advantage that we know the values of the 
couplings without having to run them. The tree level potential is just
$$
V_0(\varphi)=-{1\over 2}m^2\varphi^2+{1\over 8}\lambda\varphi^4,
$$  
with $\lambda(M_Z)$ obviously positive, so that no instability would 
arise at high $\varphi$. 
If we add one-loop corrections we discover that the dominant piece comes from
top loops [see \fig\dosl\ (a)] and is
$$
\Delta V_1=\sum_i {n_i\over 64\pi^2}M_i^4(\varphi)\left[
\log{M_i^2(\varphi)\over \mu^2}-C_i
\right]
\simeq -{12\over 64\pi^2}{1\over 4}h_t^4\varphi^4\left[
\log{h_t^2\varphi^2\over 2\mu^2}-{3\over 2}\right].
\eqn\vul
$$
[Here $n_i$ counts the number of degrees of freedom of the $i^{th}$ particle
with field-dependent mass $M_i^2=\kappa_i^2\varphi^2$ (this form for 
the masses holds for the main contributions to the potential. Higgs bosons 
have an additional field-independent piece, but their contribution is not 
important numerically). $C_i$ are some numerical constants.] We see that, 
for sufficiently large values of $\varphi$ this 
piece dominates over the tree-level part and being negative drives the 
potential to negative values. Once again, the fact that $\Delta V_1$ 
dominates over $V_0$ does not imply that the perturbative expansion is 
not well behaved. In a similar way as what we discussed in the previous 
section the expansion parameter is of the form $\alpha\log$ where again
$\alpha\sim h_t^2/16\pi^2$ and now $\log\sim\log(\phi/\mu)$. To be 
confident that the instability is really there, it has to appear for 
values of the field where $\alpha\log<1$. The fact that this can actually 
happen is then a consequence of the hierarchy  $h_t\gg \lambda$
of the relevant coupling in the loop correction to the relevant coupling in 
the tree level potential. On the other hand note that the $\log$ depends 
now on the scale $\mu$. This means that fixing $\mu\sim M_Z$ we cannot 
reliably study potential instabilities if they appear at field values 
much larger than $\varphi\sim M_Z exp(4\pi)$. Assuming this is not the case,
if $\alpha\log<1$ but not too small one should care about higher order 
corrections.
For example the dominant two-loop correction to the potential (at large 
$\varphi$) has the form  
$$
\Delta V_2(\varphi)\simeq {h_t^4\over (16\pi^2)^2} \left[
\#\left(\log{h_t^2\varphi^2\over 2\mu^2}\right)^2
+\#\left(\log{h_t^2\varphi^2\over 2\mu^2}\right)+\#
\right],
\eqn\twolv
$$ 
where the $\#$'s are field independent constants (some 
combination of couplings) [the dominant diagrams are depicted in Fig.~\dosl\ 
(b,c)]. 

\noindent{\bf Exercise} Taking advantage of the scale invariance of the 
effective potential compute the value of the constant $\#$ for the two-loop
leading log contribution  in (\use\twolv) (keeping only $h_t$ and $g_S$ 
couplings). Use the knowledge of $\Delta V_1$ and the one-loop rg functions 
$16\pi^2\gamma^{(1)}\simeq 3 h_t^2$ and $16\pi^2\beta_{h_t}^{(1)}\simeq
h_t[(9/2)h_t^2-8g_s^2]$.

\vskip .3 cm
\centerline{\epsfxsize 14. truecm \epsfbox{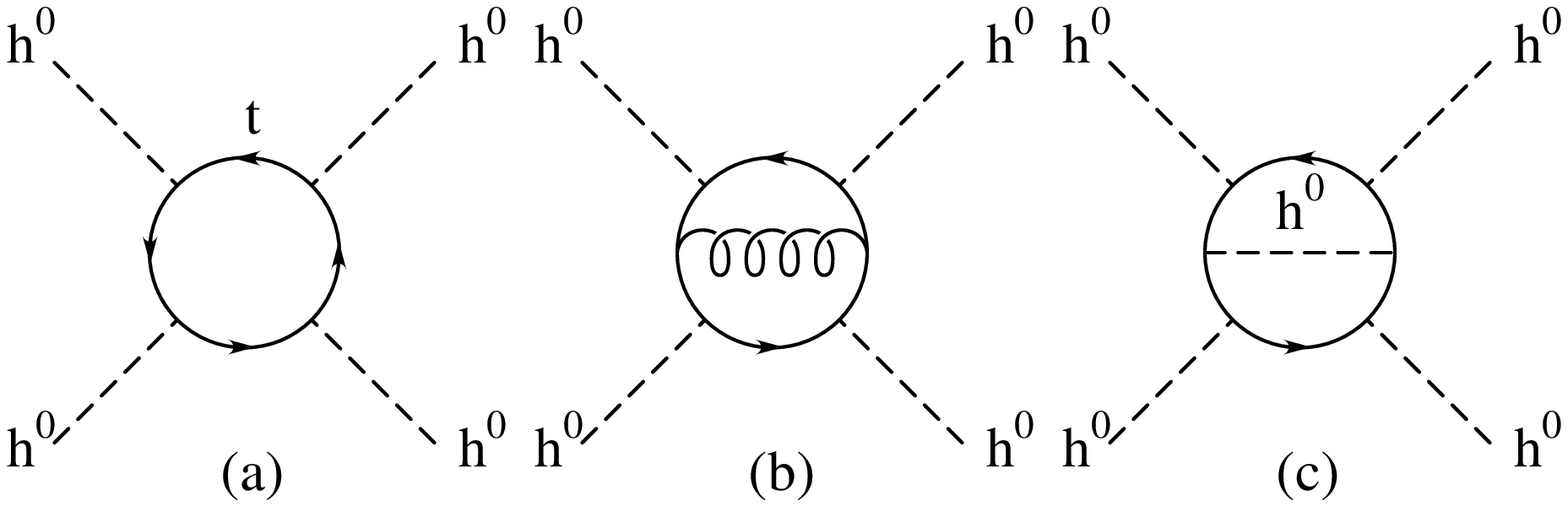} }
\nobreak
\vskip .1 cm
{\baselineskip 10 pt\narrower\smallskip\noindent\ninerm\nobreak
{ Fig.~\dosl :} Dominant loop diagrams contributing to the running of 
$\lambda$. (a) One-loop and (b,c) two-loops . \smallskip}

\vskip .4 cm

From the previous discussion it is clear that we can use our freedom in 
choosing $\mu$ to have a reliable perturbative expansion by the choice
$\mu\sim\varphi$. This choice will make small the radiative corrections
so that we can just work with the tree level potential. Of course,
this doesn't mean that the destabilization disappears (being caused by
large loop corrections). The size of radiative corrections is scale
dependent and only the full potential doesn't depend on the scale. By 
choosing $\mu\sim\varphi$ we just reorganize the expansion in such a way that
the bulk of the effect is transferred from the higher loop corrections to 
the tree level part. The instability appears then as a result of 
$\lambda(\mu\sim\varphi)<0$. We have then a clear picture of the 
procedure to follow: we consider the tree-level effective potential with 
parameters evaluated at some scale $\mu$. When $\lambda(\mu)$ runs to
negative values, it means that the potential is being destabilized at 
$\varphi$ of that order. Imposing that this destabilization scale is larger
than the cut-off scale $\Lambda$ we can compute the stability bound for that
$\Lambda$: the lowest allowed value for $\lambda(\mu\sim M_Z)$  (and thus 
for $M_H$) is such that $\lambda(\mu\sim\Lambda)=0$ (which is in practice 
used as a boundary condition for the running $\lambda$).
Implementing that program with one-loop rg functions amounts to compute the
stability bound at leading-log order.

For large $\Lambda$ (say $\Lambda\sim M_{Pl}$) the 
leading-log calculation can be a good enough approximation for the 
stability bound. A better precision can be achieved with a next-to-leading 
log calculation, proceeding  along similar lines as what was done in 
section~3. Understandably, the NTLL calculation is mandatory in
the case of lower $\Lambda$, say $1- 10\ TeV$, when the effect of NTLL terms 
can become more important.

The ingredients for a NTLL calculation should be clear by now. First, use
a one-loop effective potential with parameters running with two-loop rg 
functions. Next, keep control of the scale dependence of the potential.
This has to be done in two regions of $\varphi$: in the region near the
electroweak minimum, to ensure that the potential is minimized properly
and the Higgs mass extracted correctly (as was done in the 
previous section) and then, in the region of the instability. In that 
region one can fix 
$\mu=\alpha\varphi$ and study the flatness of the bounds with the parameter
$\alpha$. Again, it turns out that, the tree-level calculation of the
bound requires a careful choice of the scale while the one-loop calculation
is much more stable. 

As a result of using a one-loop expression for the potential, one should
replace the boundary condition $\lambda(\Lambda)=0$ by a one-loop corrected
version of it ${\widetilde \lambda}(\Lambda)=0$. Here ${\widetilde \lambda}$
is basically the one-loop coefficient of $\varphi^4$ 
(once $\mu=\alpha\varphi$). From (\use\vul)
$$
{\widetilde \lambda}\simeq\lambda + \sum_i {n_i\over 
8\pi^2}\kappa_i^2\left[ \log{\kappa_i^2\over \alpha^2}-C_i
\right].
$$
The fact that ${\widetilde \lambda}$ should be used for a faithful 
description of
the potential destabilization is exemplified by \fig\destab\ . For the 
indicated values of the parameters, the first plot gives the behaviour of 
the 
potential. Note the appeareance of a deep non-standard minimum at large 
values of the field. The second is a plot of the running of $\lambda$ and 
${\widetilde \lambda}$ with the scale, showing that the scale of 
destabilization of the potential indeed coincides with the scale at which 
${\widetilde \lambda}$ crosses zero. The point were $\lambda$ is zero
is instead an order of magnitude lower.

\centerline{\epsfxsize 15. truecm \epsfysize 9. truecm \epsfbox{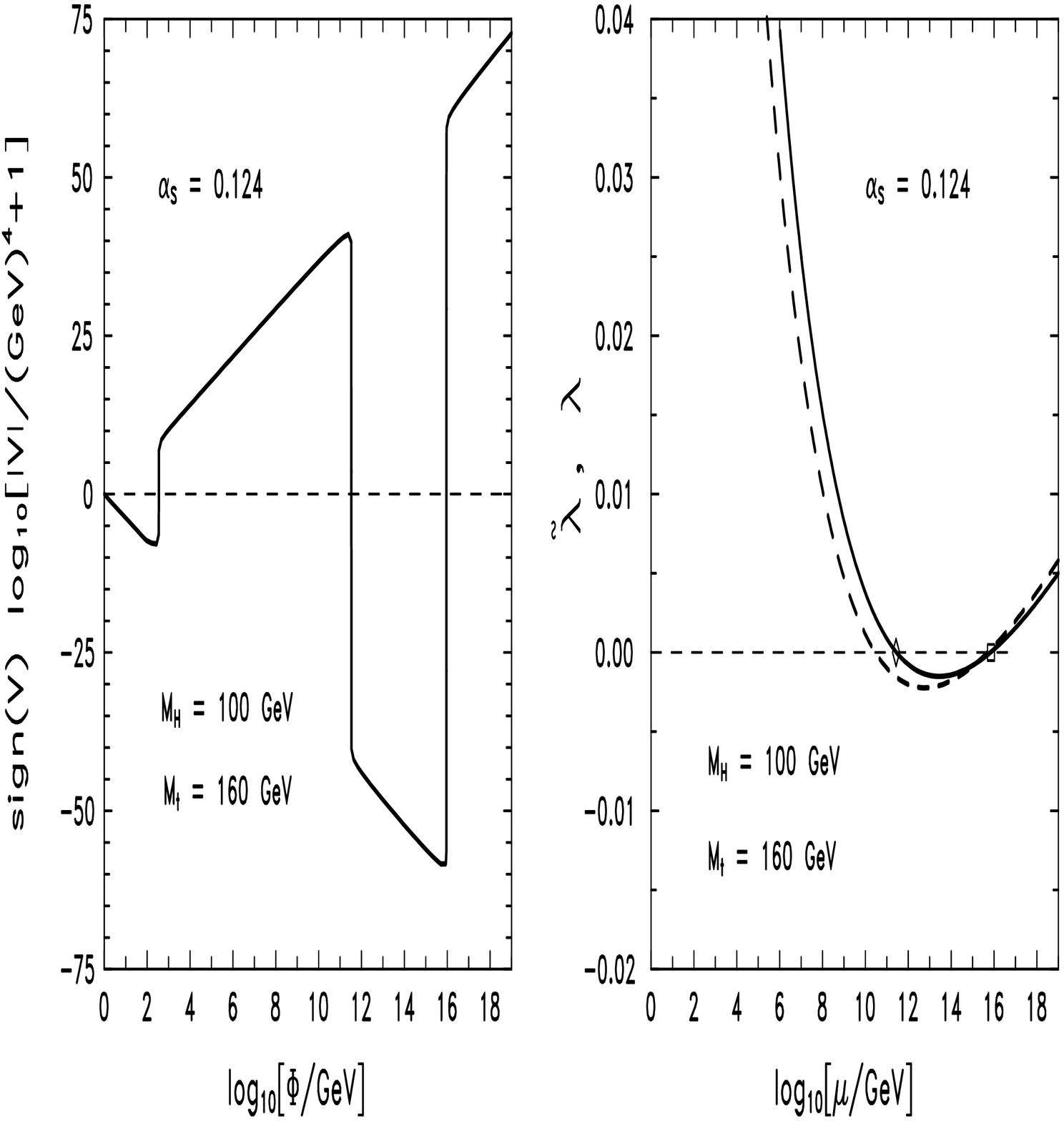} }
\nobreak
\vskip .1 cm
{\baselineskip 10 pt\narrower\smallskip\noindent\ninerm\nobreak
{ Fig.~\destab : Right: Effective potential. Left: Running $\lambda$ 
(dashed) and ${\widetilde \lambda}$ (solid).}  \smallskip}

\vskip .4 cm

Note that the potential does not run to $-\infty$ for large field values 
but rather turns again to positive values. This is caused by the 
decrease of $h_t$ at high scales. The destabilizating force then weakens
and eventually the effect of gauge couplings takes over making the couplings
$\lambda$ and ${\widetilde \lambda}$ positive again (see left plot in 
Fig.~\destab\ ).

The last ingredient for the NTLL computation of the bound is to  
extract physical masses for $M_H$ and $M_t$ as explained in 
Section~\MSSM\ [Eqs. (\use\physh) and (\use\physt)].
The NTLL results are typically ${\cal O}(10)\ GeV$ lower than the LL ones
and even more [${\cal O}(20)\ GeV$] for low cut-offs [e.g. $\Lambda\sim 
10^3\ GeV$]. The bounds, as a function of $M_t$ and $\Lambda$ are plotted
in \fig\stabound\ . There is a non-negligible dependence on $\alpha_s$
(larger for larger $\Lambda$) the bound being larger for
smaller $\alpha_s$. 

\centerline{\epsfxsize 15. truecm \epsfysize 9. truecm 
\epsfbox{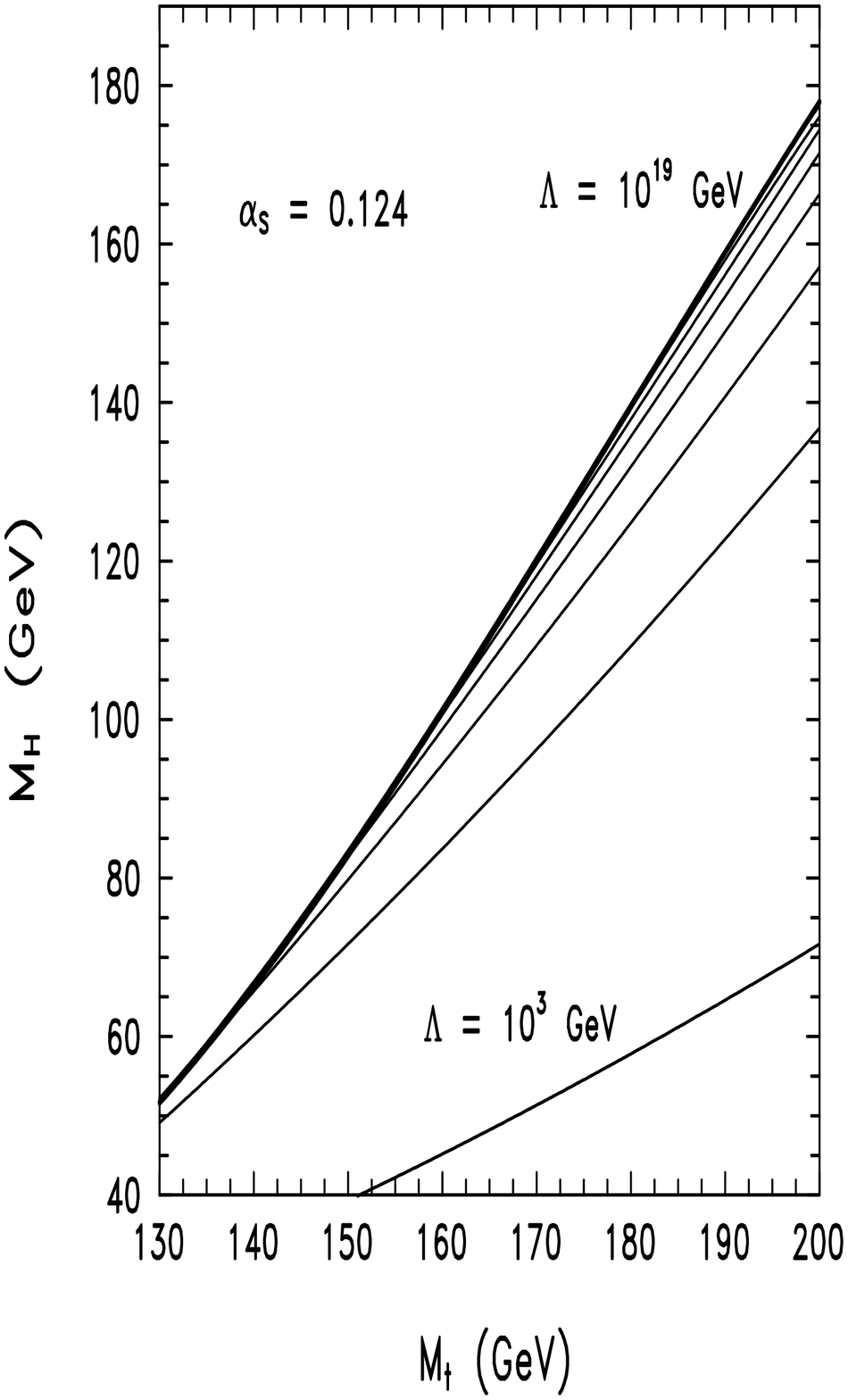} } \nobreak
\vskip .1 cm
{\baselineskip 10 pt\narrower\smallskip\noindent\ninerm\nobreak
{ Fig.~\stabound : Stability bounds for the Standard Model Higgs mass
as a function of the top mass $M_t$ and the cut-off scale $\Lambda$ from
$\Lambda=10^3\ GeV$ to $10^{19}\ GeV$ in steps of two orders of 
magnitude. }  \smallskip}

\vskip .4 cm

The area below a given curve corresponds to a region of 
parameters where the effective potential develops an instability at field 
values below the corresponding cut-off. This cut-off is usually referred 
to as the scale of new physics and should be interpreted with some care.
The correct way of thinking about it is that, for a given pair 
$(M_t,M_h)$ in the instability area, there is a curve labeled by some 
value of $\Lambda$ that goes through that point. That value of $\Lambda$ 
is the scale at which the instability develops in the pure Standard Model
for the specified values of $M_t$ and $M_h$. As such, it can be computed
very precisely, irrespective of whether $\Lambda$ is large or small.
In case that for the experimental value $(M_t,M_h)$ the instability scale
$\Lambda(M_t,M_h)$ is below the Planck mass it 
would imply that the Standard Model would have to be extended in such a 
way that the instability is cured. 
This means that new physics should have an effect at 
scales below $\Lambda$. 

\section{Standard Model Metastability Bound on the Higgs Mass}
\tagsection\Metastab

Strictly speaking, we may admit the possibility that the electroweak
vacuum is not the deepest vacuum of the effective potential provided it
is sufficiently long lived and the decay probability to the true
vacuum is negligible. In the presence of a large potential barrier
separating the two vacua it is reasonable to expect that the decay rate
for tunneling (per unit time and unit volume) would be exponentially 
suppressed. However
the Universe is old and large and it turns out that for some choices of the
parameters below the stability curve ( e.g. for $M_t=180\ GeV$ and $M_h=100\ 
GeV$) the unstable electroweak vacuum should have decayed long ago and are 
thus unacceptable if no new physics is at work at (or below) the scale of 
the instability. Then, the requirement that an unstable vacuum is
acceptable if its lifetime is longer than the age of the Universe weakens 
the stability bounds derived in the previous section increasing the allowed
range of Higgs masses. 

However, there is a more stringent requirement an 
unstable electroweak vacuum should meet: it has to survive the high 
temperatures in the early Universe, when thermal excitations can trigger the
decay to the true vacuum. The Higgs potential in the presence of the
thermal plasma in the early Universe depends strongly on the temperature, 
$T$.   
For $T$ much larger that the scale of instability, $\Lambda$, 
$SU(2)\times U(1)$ symmetry is restored and the only minimum of the 
potential is at 
the origin. At $T\sim\Lambda$ a new local minimum 
appears at a scale $\phi\sim T\sim \Lambda$ and at some temperature 
$T_c^1$ 
the new minimum becomes degenerate with the one at the origin. For $T<T_c^1$
the minimum away from the origin is the deepest and the decay from the 
minimum at the origin to it becomes possible. For lower and lower 
temperatures the true minimum gets deeper and deeper evolving to the 
non-standard $T=0$ vacuum at $T\ll \Lambda$. The barrier between the two 
vacua is always present. Eventually, if the Universe remains trapped at the
origin, at a temperature $T_c^{EW}$ of order $100\ GeV$ the standard 
electroweak 
phase transition takes place, $SU(2)\times U(1)$ gets broken and the
Universe sits in a new minimum that lies at the electroweak scale.
The possibility of a no-return transition to the 
non-standard minimum is more likely than tunneling from quantum 
fluctuations at $T=0$. For example, if $M_t=180\ GeV$ and $M_h=130\ GeV$,
the unstable electroweak vacuum would be safe against quantum tunneling
but cannot survive the high temperature early Universe. 

The phase transition to the non-standard minimum is strongly first order 
and proceeds via thermal nucleation of bubbles of true phase that grow till 
they fill the Universe. If a bubble of true phase is too small it collapses
under surface tension while, if it is large enough the gain in potential 
energy compensates surface tension and the bubble grows. There is then a
critical bubble, a saddle point of the free energy functional, that gives
the critical energy that appears in the Boltzmann exponent for the
decay rate. Consider a static and spherically 
symmetric bubble of true vacuum with Higgs profile $\phi(r)$. Here $r$ is 
spatial distance from the center of the bubble, so that $\phi(0)$ probes the
instability region of the potential and $\phi(r\rightarrow\infty)$ goes 
to the false vacuum. At a given temperature, the extra energy for such a 
configuration in a sea of false vacuum is given by 
$$
E[\phi(r)]=4\pi\int_0^\infty r^2 dr \left[
{1\over 2}\left({d\phi\over dr}
\right)^2+V(\phi,T)-V(v_{false},T)
\right].
$$
The critical bubble minimizes this energy and so the critical 
bubble profile is the  solution of the Lagrange equation
$$
{d^2\phi\over dr^2}+{2\over r}{d\phi\over dr}={dV\over d\phi},
$$
with boundary conditions
$$
\phi(r\rightarrow\infty)=v_{false}(T),\;\;\;\;\;\;\;\;\;
d\phi/ dr |_{r=0}=0.
$$
Here $v_{false}(T)$ is the field expectation value in the minimum close
to the origin: $v_{false}=0$ for $T>T_c^{EW}$ and $v_{false}=v_{EW}(T)$
for $T<T_c^{EW}$.
For a given temperature T, the effective potential is known and the
critical bubble profile $\phi_B(r)$ and energy $E_B(T)=E[\phi_B(r)]$ can be  
computed. The rate of nucleation of these bubbles per unit time and unit 
volume is then given by 
$$
\Gamma/\nu\sim \omega T^4 \exp{[-E_B(T)/T]},
\eqn\rate
$$
with $\omega$ a constant that can be taken of order unity for our purpose.
The qualitative dependence of $E_B$ with temperature is shown in 
\fig\ecrit\ . For Higgs masses larger than the LEP I experimental bound 
$\sim 65\ GeV$ the deepest minimum of the curve appears always in the region
$T\gg T_c^{EW}$ so that the decay rate is maximal at temperatures much larger
than $T_c^{EW}$.

To get the probability of decay, the rate (\use\rate) should be 
multiplied by the volume of our current horizon scaled back to 
temperature $T$. The differential probability is then
$$
{dP\over d\log T}=\kappa {M_{Pl}\over T}\exp{(-E_B/T)},
$$
with $\kappa\sim 3.25\times 10^{86}$ and is plotted also in fig.~10 . 
Obviously it peaks near the minimum of $E_B/T$. The integrated probability
of decay is then 
$$
P=\int_0^{T_c^1}{dP(T')\over dT'}dT'.
\eqn\intp
$$
Note that this probability is not normalized to one. In fact its 
interpretation is that the fraction of space that remains in the old 
metastable phase is given by $exp(-P)$. So, $P\gg 1$ (see \fig\logp ) 
indicates that 
the metastable electroweak vacuum would have decayed to the non-standard 
minimum in the hot early epoch of the Universe.

\centerline{\epsfxsize 15. truecm \epsfysize 9. truecm \epsfbox{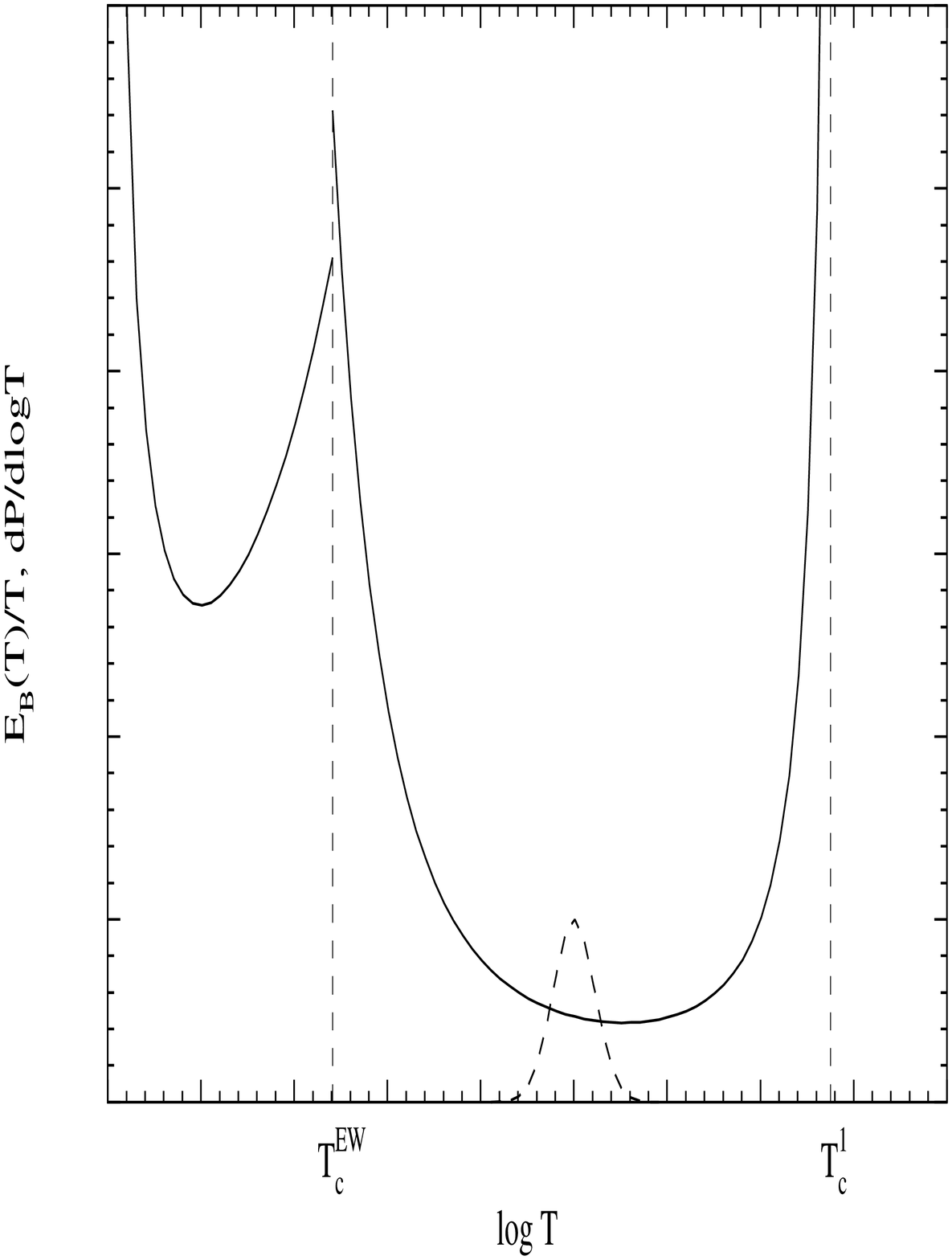} }
\nobreak
\vskip .1 cm
{\baselineskip 10 pt\narrower\smallskip\noindent\ninerm\nobreak
{ Fig.~\ecrit : Ratio of the critical bubble energy to the temperature 
(solid line) 
and differential probability for bubble nucleation (dashed) as a function of 
temperature. $T_c^{EW}$ and $T_c^1$ are the two critical 
temperatures.}  \smallskip}

\centerline{\epsfxsize 15. truecm \epsfysize 9. truecm 
\epsfbox{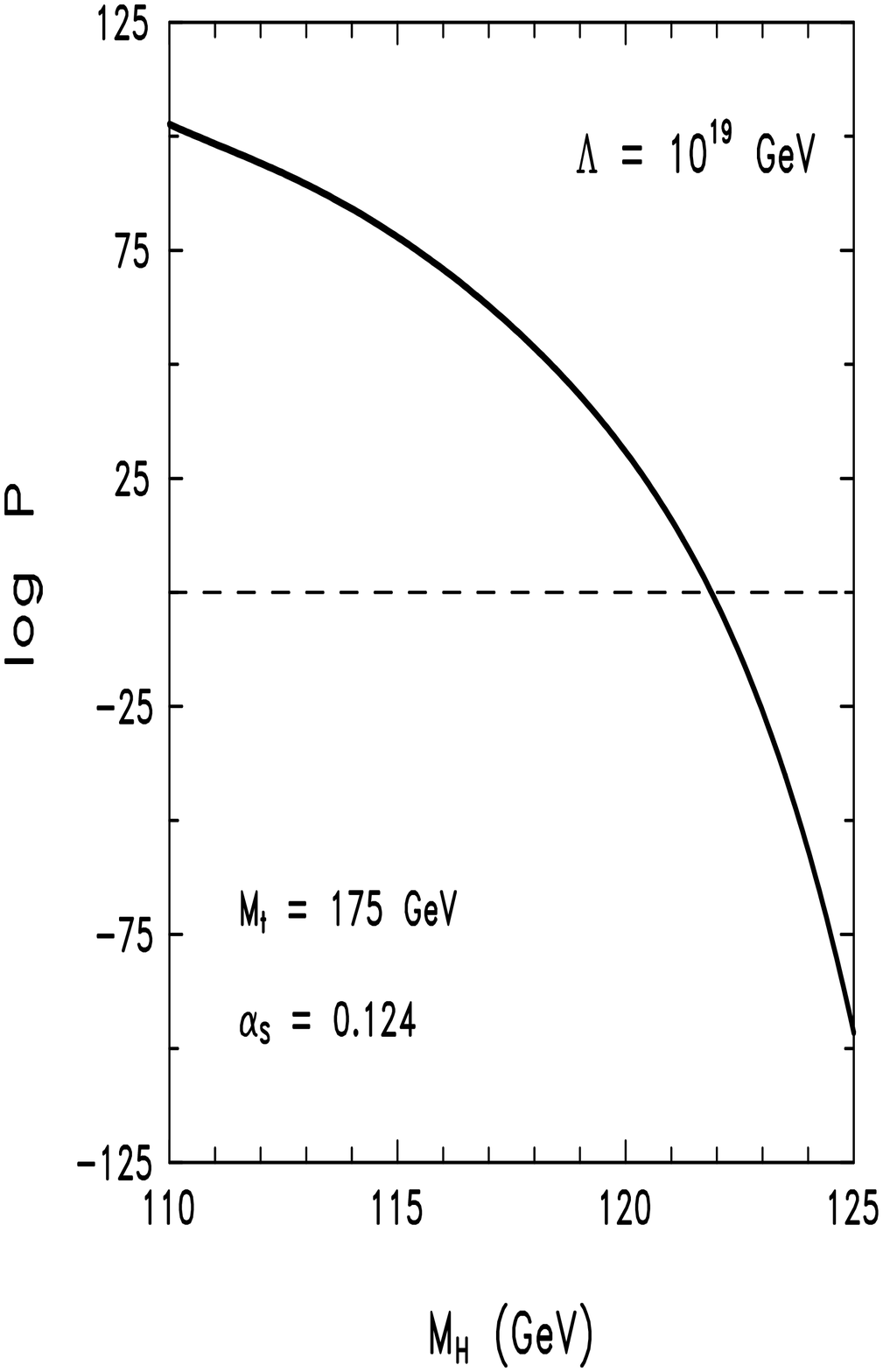} } \nobreak
\vskip .1 cm
{\baselineskip 10 pt\narrower\smallskip\noindent\ninerm\nobreak
{ Fig.~\logp : Logarithm of the total probability of decay to the true 
vacuum as a function of $M_h$ for $M_t=175\ GeV$. The point where the 
curve roughly crosses the dashed line gives the metastability bound.}  
\smallskip}

\vskip .4 cm

Then, for fixed $M_t$ the probability $P$ of thermal decay to
the true vacuum can be computed as a function of $M_h$. One example is 
shown in fig.~\use\logp\ for $M_t=175\ GeV$. Note the vast range of 
variation 
of $P$ when $M_h$ is changed by a few $GeV$. Actually this is due to the 
exponential sensitivity of the decay rate to $E_B$ and then to the shape of
the effective potential. From fig.~\use\logp\ we can derive the so 
called  metastability bound for $M_t=175\ GeV$: below $\sim 122\ GeV$
$P$ is very large and the vacuum decays quickly while for values above that,
$P$ is exponentially small and the  electroweak vacuum, if metastable, is 
sufficiently long lived. The critical value $P=1$ can be taken to compute 
the critical mass but as is clear from the figure the 
bound is insensitive to the exact value chosen for the 
critical probability [provided is very roughly ${\cal O}(1)$].

The metastability bounds would depend also on the cut-off scale $\Lambda$ 
where new physics is expected to appear\footnote{$^\dagger$}{Again the 
precise meaning of this scale is provided by the calculation itself as 
the scale at which the effect of new physics should affect the finite 
temperature potential in a way suitable of modifying the decay rates.}. 
By definition we can compute reliably the effective potential at field 
values  lower than $\Lambda$ and temperatures lower than 
$T_\Lambda\sim\Lambda$. 
The integrated probability that should be required to be less than ${\cal 
O}(1)$ is then (\use\intp) with this temperature cut-off implemented
$$
P(T_\Lambda)=\int_0^{T_\Lambda}dP(T').
$$
For some fixed values of $M_t$ and $M_h$, the condition $P(T_\Lambda)=1$
gives $T_\Lambda$. The profile of the critical bubble at that temperature 
can be calculated and in particular the value of the field at the center 
of the bubble $\phi_B(0)$ obtained. To avoid the decay of the electroweak 
vacuum for this choice of parameters, new physics should modify the 
effective potential at values of the field of the order $\phi_B(0)$, i.e.
new physics should appear at $\Lambda=\phi_B(0)$.

The numerical results for the metastability bounds are shown in \fig\metab\
for two values of $\Lambda$: $10^{19}\ GeV$ (three upper curves) and $10\ 
TeV$ (three lower curves). For a given $\Lambda$ the $(M_t,M_h)$ plane gets 
divided in four different regions. Above the long dashed line the electroweak
vacuum is absolutely stable, while below it is only metastable: there is 
a deeper non-standard minimum developing below $\Lambda$. Between the 
absolute 
stability lower bound and the solid line the metastable electroweak 
vacuum is long lived and acceptable. Below the solid line however, it 
would have decayed by thermal excitations in the early Universe. The 
short dashed line indicates the metastability bound for quantum tunneling 
at zero T. For $\Lambda=10^{19}\ GeV$ the lower metastability bound on 
the Higgs mass is (for $\alpha_s=0.118$) 
$$
M_h(GeV)>2.306[M_t(GeV)-180]+138.
$$
The implications of these bounds for the existence of new physics beyond 
the Standard Model are discussed in the next and last section.

\centerline{\epsfxsize 15. truecm \epsfysize 9. truecm 
\epsfbox{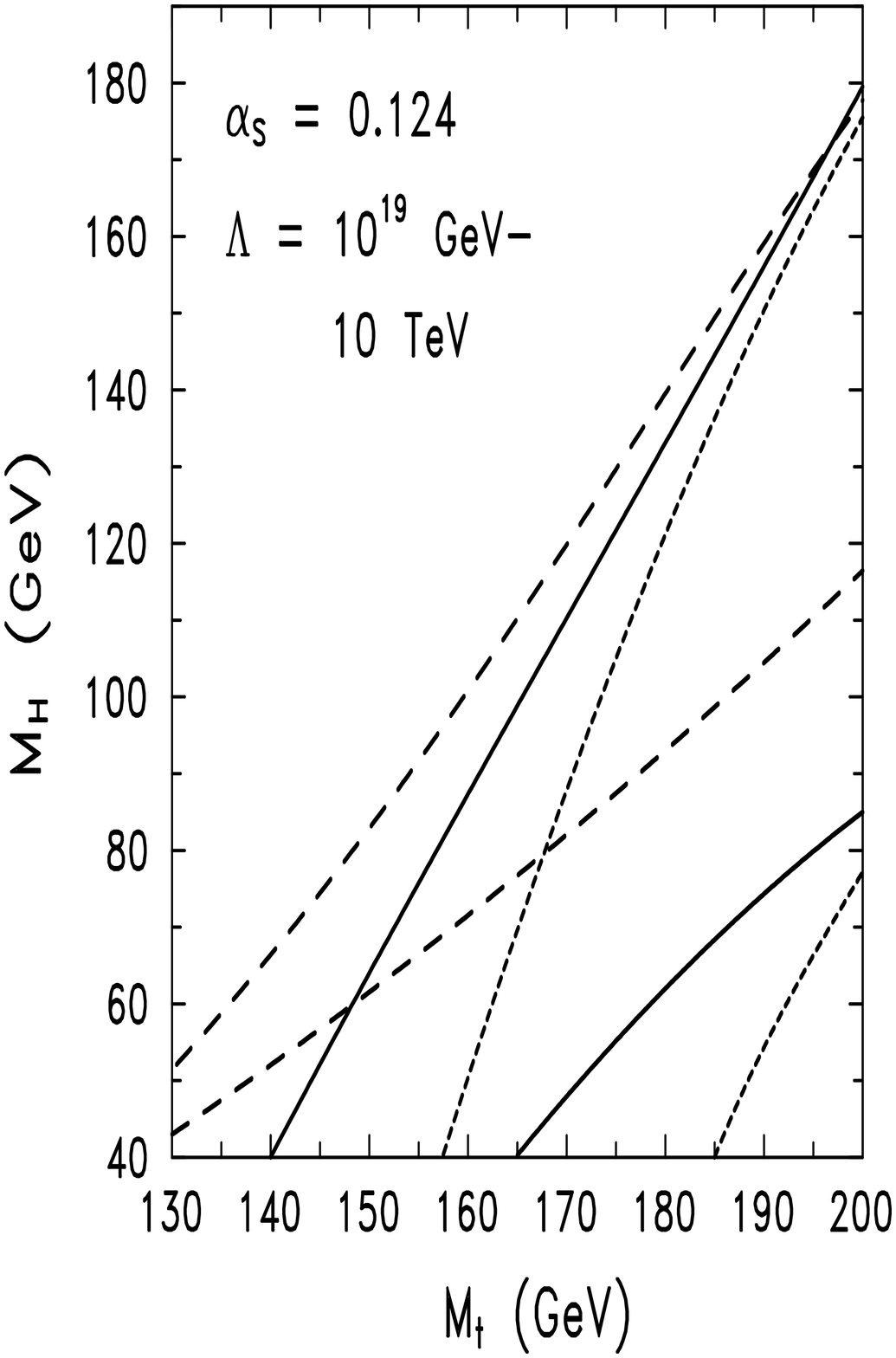} } \nobreak
\vskip .1 cm
{\baselineskip 10 pt\narrower\smallskip\noindent\ninerm\nobreak
{ Fig.~\metab : Lower bounds on the Higgs mass for $\Lambda=10^{19}\ 
GeV$ (upper set) and $\Lambda=10\ TeV$ (lower set). Long dashed curves 
give the absolute stability bound, short dashed the metastability bound 
for quantum tunneling and solid for thermal tunneling in the hot early 
Universe.}  \smallskip}

\vskip .4 cm

\section{Implications}
\tagsection\Impl

The most important implication of the {\it lower} bounds on the Higgs mass 
derived in section 5 is that the measurement of $M_h$ may provide
an {\it upper} bound on the scale $\Lambda$ of new physics. For this to be
true, the top mass should be heavy enough so that some instability would
appear in the Standard Model potential if the new physics were not present.
Numerically the requirement is (all masses in $GeV$)
$$
M_t > {M_h\over 2.25} + 123\, GeV.
\eqn\heavy
$$
(This is obtained from the metastability bound for $\Lambda=10^{19}\ GeV$).
In particular, for Higgs masses above the experimental limit $\sim 65\ GeV$
the top quark should be heavier than $\sim 152\ GeV$, which is the case.
Similarly, (\use\heavy) also tells that, unless $M_t< 160\ GeV$, the 
discovery of a Higgs boson by LEP II would imply that the Standard Model
cannot be valid up to the Planck Scale.

Moreover, in most of the parameter space, the scale below which new 
physics should enter is much smaller than $M_{Pl}$. Consider for example
\fig\impli\  where metastability bounds for different top masses are plotted
as a function of the scale $\Lambda$. Suppose now that LEP II finds a 
$90\ GeV$ Higgs boson. Then, if $M_t>170\ GeV$, we can read from the figure
that new physics should appear below $\Lambda\sim10^6\ GeV$. Otherwise, the
Standard Model potential would be unstable and our vacuum would have decayed
long ago.

\centerline{\epsfxsize 15. truecm \epsfysize 9. truecm 
\epsfbox{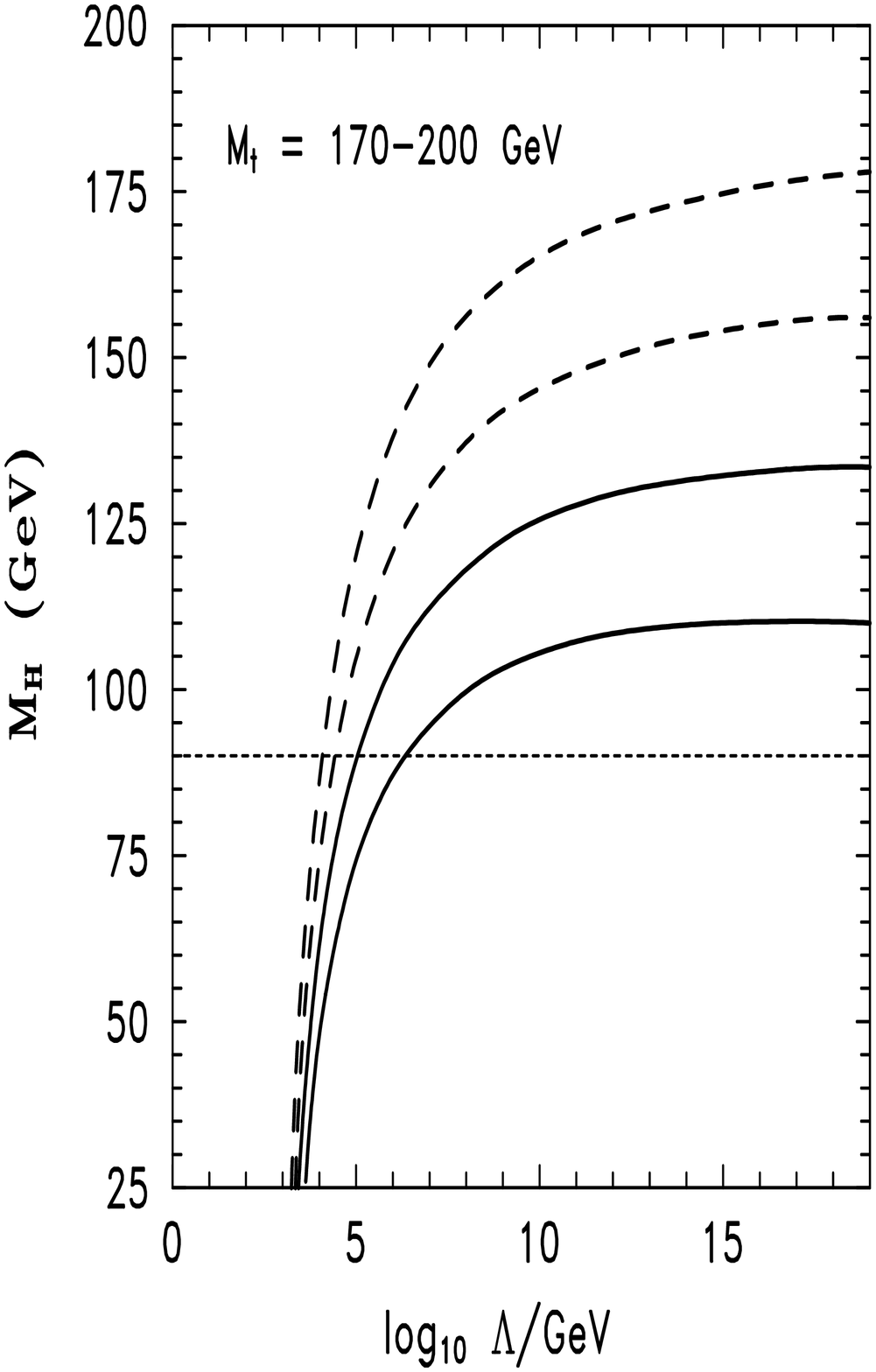} } \nobreak
\vskip .1 cm
{\baselineskip 10 pt\narrower\smallskip\noindent\ninerm\nobreak
{ Fig.~\impli : Metastability bounds on the SM Higgs boson mass as a 
function of the cut-off $\Lambda$ for the indicated values of the top 
mass (lower curve for lower $M_t$). The horizontal line indicates roughly 
the reach of LEP II.}  \smallskip}

\vskip .4 cm

Of course, the new physics should be such that this instability is cured.
For example, a heavy fourth generation would make the effective potential
even more unstable (a new large Yukawa coupling would add to the 
destabilization effect of the top quark). Then, this kind of 
stability analysis  
can in fact be used to constrain such extensions of the Standard Model
(see A. Novikov's contribution to these proceedings).
As a prime example of new physics that would go in the good direction we
can consider the Minimal Supersymmetric Standard Model. In the MSSM, it turns
out that the running of the quartic Higgs coupling is no longer dominated
by the top Yukawa coupling. The effect of top quark loops is compensated by
stops, the supersymmetric partners of the top. The relevant diagrams are 
shown in \fig\superi\ . In fact, quartic Higgs couplings in the MSSM run 
like gauge couplings. This effect is schematically depicted in 
fig.~\superi\ . Below the supersymmetric 
scale $\Lambda$ the quartic Higgs coupling decreases steeply due to top loop
effects. But once the supersymmetric threshold is crossed, all supersymmetric
particles influence the running of $\lambda$ with the effect of Yukawas
cancelling and leaving just  the gauge renormalization. The figure then shows
how the presence of the supersymmetric threshold prevents $\lambda$ to 
become negative and makes it rise again, thus stabilizing the potential.

In the framework of the MSSM as model for new physics, it is then 
tempting to try and put an upper bound on the supersymmetric scale by using
metastability arguments. Unfortunately, as is shown already in 
fig.~\impli\ , metastability bounds cannot compete with simple naturalness
criteria that require $\Lambda_{SUSY}< {\cal O} (1)\ TeV$.

\centerline{\epsfxsize 15. truecm \epsfysize 9. truecm 
\epsfbox{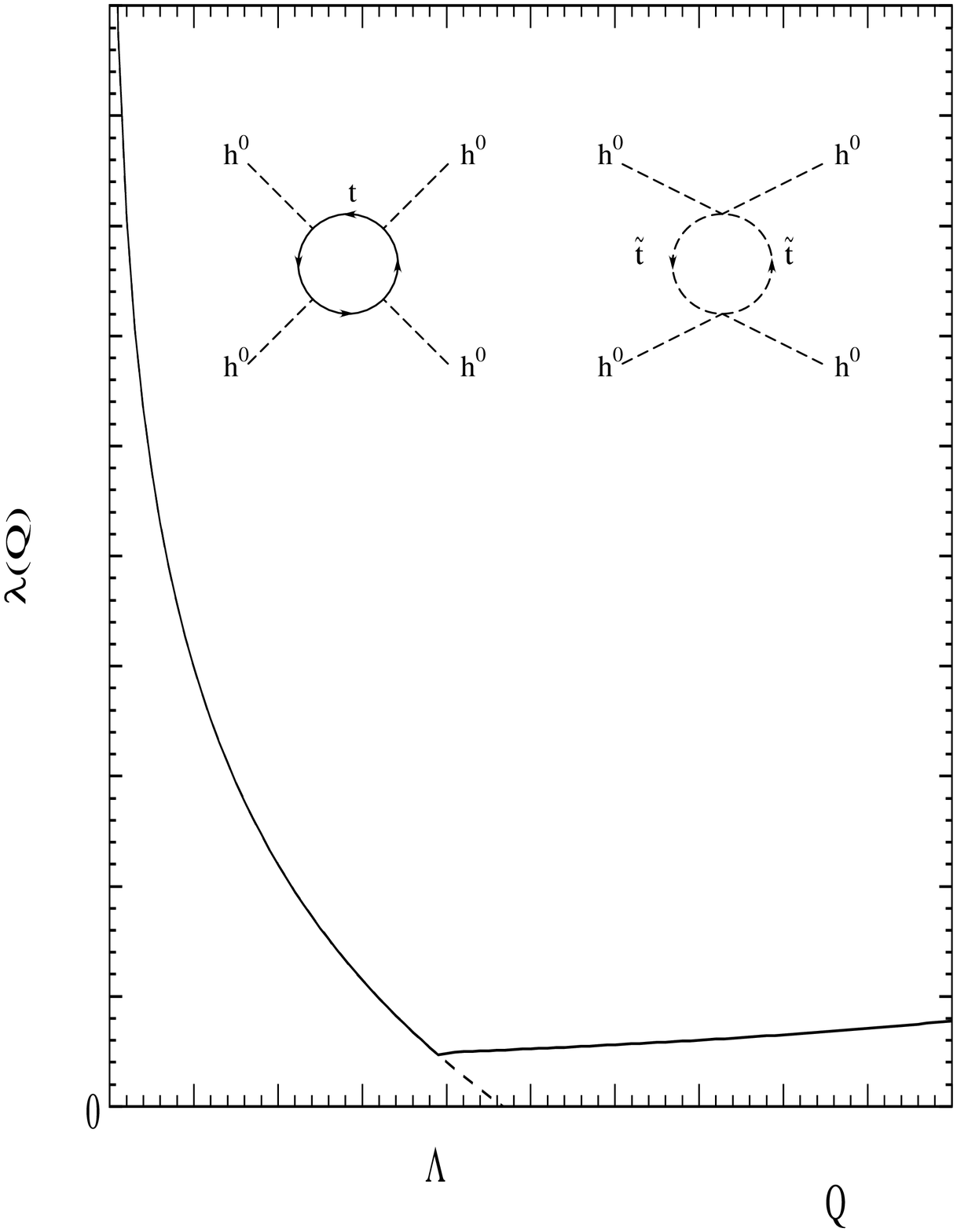} } \nobreak
\vskip .1 cm
{\baselineskip 10 pt\narrower\smallskip\noindent\ninerm\nobreak
{ Fig.~\superi : Schematic running of the quartic Higgs coupling below and 
above the supersymmetric threshold with the relevant diagramas 
indicated. The pair $(M_t,M_h)$ is 
chosen to lie in the instability region.}  \smallskip}

\vskip .4 cm

Nevertheless, useful information can be extracted from the confrontation 
of the theoretical expectations for Higgs masses in the Standard Model and
the MSSM. In \fig\smmssm\ we plot (steepest curve) the less restrictive 
metastability bound for the SM Higgs mass, choosing $\Lambda=10^{19}\ 
GeV$, i.e. assuming validity of the model up to the Planck scale. 
Superimposed in the same plot it is shown the absolute upper bound on the 
lightest Higgs mass in the MSSM for $\Lambda_{SUSY}=1\ TeV$ (more or less 
the limit from naturalness). In both curves the uncertainty caused by 
$\alpha_S$ in the range indicated is reflected in the dashed curves.

The $(M_t,M_h)$ plane gets divided into four different zones. The 
experimental determination of the zone actually realized is of great 
interest. If the Higgs boson is found in the upper left region it would
be compatible with the SM but it would be too heavy for the MSSM (remember
that we are talking about a neutral Higgs boson with properties similar 
to the SM one). Alternatively, if it is the lower right area the one chosen
by nature, then the Higgs mass would be too low for the pure SM and new 
physics should appear below the Planck mass. This new physics could well be
in the form of the MSSM, and in that region the mass of the Higgs is 
compatible with that hypothesis. Finally, the region in the lower left 
corner is compatible with the SM and MSSM while the upper right
corner is incompatible with both. 

\centerline{\epsfxsize 15. truecm \epsfysize 9. truecm 
\epsfbox{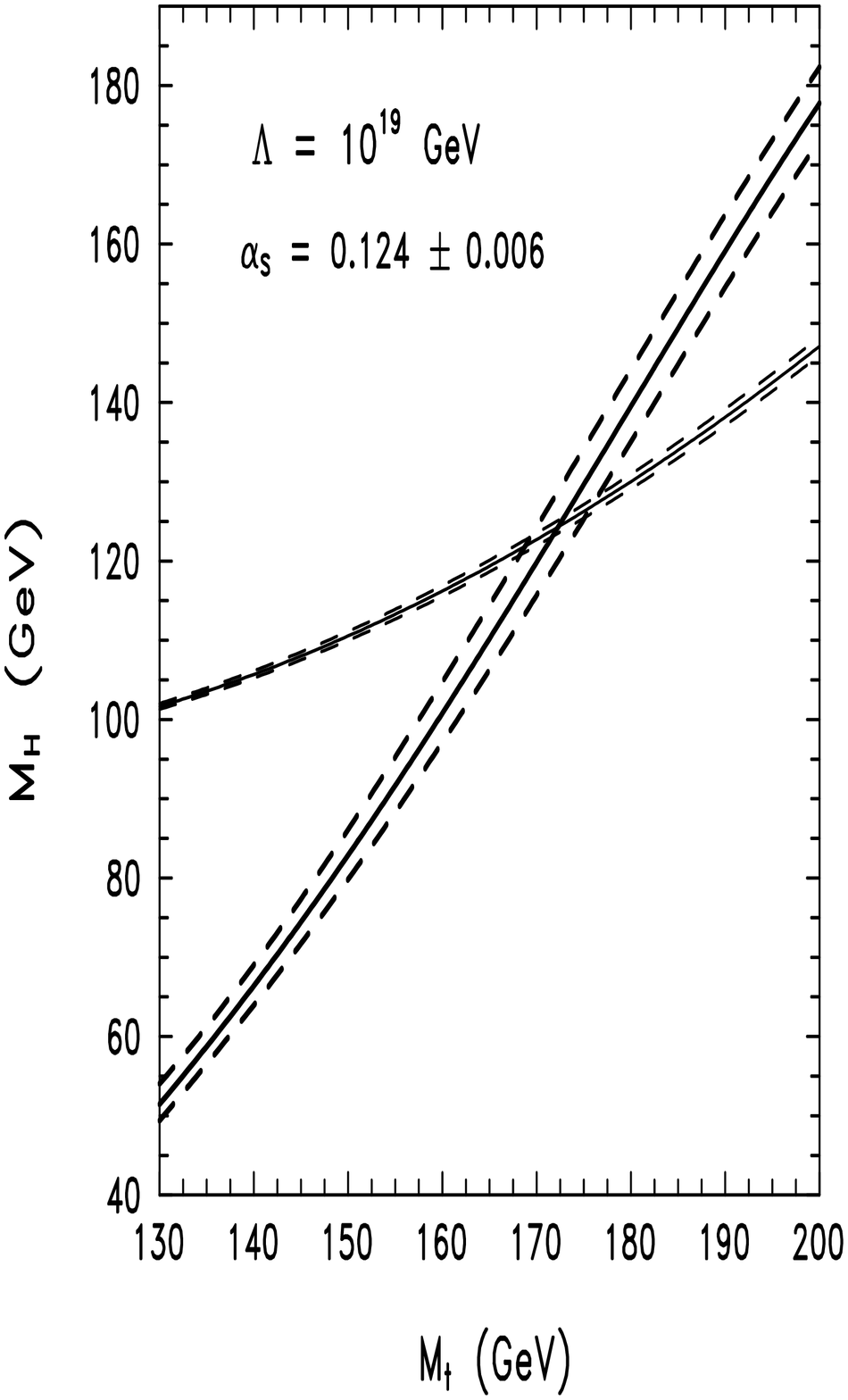} } \nobreak
\vskip .1 cm
{\baselineskip 10 pt\narrower\smallskip\noindent\ninerm\nobreak
{ Fig.~\smmssm : Absolute lower metastability bound on the SM Higgs mass
with $\Lambda=10^{19}\ GeV$ (steepest curves) and absolute upper bound on 
the lightest Higgs mass in the MSSM with $M_S=1\ TeV$ and maximal stop 
mixing effect. Dashed lines show the variation when $\alpha_s$ is changed 
in the indicated range.}  \smallskip}

\vskip .4 cm

So, at least in some regions of parameter space, we may be able to 
distinguish indirectly between the lightest MSSM Higss boson in the 
decoupling SUSY limit and the SM Higgs, which is a very difficult task 
experimentally.
In any case it is remarkable that the detection of a Higgs boson, the 
last missing ingredient of the Standard Model, can already provide 
information against the pure SM and point towards the need of new physics
at some scale well below the Planck scale.

 \vskip .5 cm

I would like to thank Marcela Carena, Alberto Casas, Denis Comelli,
Mariano Quir\'os, Antonio Riotto and Carlos Wagner for 
fruitful and enjoyable collaborations on some of the topics covered here.
I also express my gratitude to the organizers of the school for their kind 
invitation to lecture and their warm hospitality and to all 
the participants for contributing to a most charming atmosphere. This work 
was supported by the Alexander-von-Humboldt Foundation.

\vfill
\section{References}
\tagsection\Ref
\noindent
{\it \Ref.1 Section 2}
\vskip .1 cm
 
The bounds presented in Section 2 were derived in the 
following references:

\noindent$\bullet$ Langacker, P. and Weldon, H.A., {\it Phys. Rev. Lett.} 
{\bf 
52}, 1377 (1984); Weldon, H.A., {\it Phys. Lett.} {\bf B146}, 59 (1984).

\noindent$\bullet$ Comelli, D. and Espinosa, J.R., DESY Preprint to appear.

For analytical and numerical bounds on the lightest Higgs boson in 
general supersymmetric models see:

\noindent$\bullet$ 
Espinosa, J.R. and Quir\'os, M., {\it Phys. Lett.} {\bf B302}, 
51 (1993); {\it Phys. Lett.} {\bf B279}, 92 (1992); Espinosa, J.R.,
{\it Phys. Lett.}{\bf B353}, 243 (1995);
Kane, G., 
Kolda, C. and Wells, J.D., {\it Phys. Rev. Lett.} {\bf 70}, 2686 (1993).

A nice example of the decoupling limit in multidoublet Higgs 
models, with all Higgses heavy but one `true' light Higgs boson, is studied in

\noindent$\bullet$ Georgi, H. and Nanopoulos, D.V., {\it Phys. Lett.} {\bf 
82B}, 95 (1979).
  
For a rather complete analysis of the decoupling limit  in multi 
Higgs model see

\noindent$\bullet$ Haber, H.E. and Nir, Y., {\it Nucl. Phys.} {\bf B335}, 363 
(1990).

The analysis of Higgs production in the presence of gauge 
singlets is based on

\noindent$\bullet$ Kamoshita, J., Okada, Y. and Tanaka, M., 
 {\it Phys. Lett.} {B328}, 67 (1994).
\vskip .2 cm

\noindent
{\it \Ref.2 Section 3}
\vskip .1 cm
For an introduction to the MSSM see e.g. Kazakov's lectures at this school. 

The important effect of radiative corrections in the mass of the lightest 
Higgs boson in the MSSM was put forward in 

\noindent$\bullet$ 
Okada, Y., Yamaguchi, M. and Yanagida, T., {it Prog. Theor. Phys.}
{\bf 85}, 1 (1991); {\it Phys. Lett.} {\bf B262}, 54 (1991);
Haber, H.E. and Hempfling, R. {\it Phys. Rev. Lett.} {\bf 66}, 1815 (1991);
Ellis, J., Ridolfi, G., and Zwirner, F., {\it Phys. Lett.} {\bf B257} 83 
(1991); {\it ibid.} {\bf B262}, 477 (1991); Barbieri, R., Frigeni, 
M. and Caravaglios, F., {\it ibid.} {\bf B258}, 167 (1991).

The NTLL calculation of the Higgs boson mass described in 
the text is in

\noindent$\bullet$ 
Casas, J.A., Espinosa, J.R., Quir\'os, M. and Riotto, A. {\it 
Nucl. Phys.} {\bf B355}, 3 (1995).

Complimentary work can be found in

\noindent$\bullet$ 
Hempfling, R. and Hoang, H., {\it Phys. Lett.} {\bf 
B331}, 99 
(1994); Kodaira, J., Yasui, Y. and Sasaki, K., {\it Phys. Rev.} {\bf 
D50}, 7035 (1994), Haber, H.E., Hempfling, R. and Hoang, H., Preprint 
CERN-TH/95-216 to appear.

For a good introduction to the use of effective theories see 

\noindent$\bullet$ Cohen, A.G., in Proceedings of TASI-93,  
World Scientific, 1994.

Analytical formulas for the MSSM Higgs spectrum that include the most 
important loop corrections are worked out in

\noindent$\bullet$ 
Carena, M., Espinosa, J.R., Quir\'os, M. and Wagner, C., {\it 
Phys. Lett. } {\bf B355}, 209 (1995); 
Carena, M., Quir\'os, M. and Wagner, C., {\it Nucl. Phys.} {\bf B461}, 
407 (1996).

The effective theory below the SUSY scale in the case of low $m_A$ is 
studied (with particular attention to the mass of 
the lightest Higgs) in the nice paper

\noindent$\bullet$ 
Haber, H.E. and  Hempfling, R., {\it Phys. Rev.} {\bf D48}, 4280 
(1993).
\vskip .2 cm

\noindent
{\it \Ref.3 Section 4}
\vskip .1 cm
For a review on the Higgs potential, see

\noindent$\bullet$ Sher, M., {\it Phys. Rep.} {\bf 179}, 274 (1989).

Lower stability bounds on the SM Higgs mass were originally studied in

\noindent$\bullet$ Krasnikov, N.V. {\it Yadern. Fiz.} {\bf 28}, 549 (1978);
Politzer, H.D. and Wolfram, S., {\it Phys. Lett.} {\bf 82B}, 242 (1979);
Hung, P.Q., {\it Phys. Rev. Lett.}{\bf 42}, 873 (1979);
Cabibbo, N., Maiani, L., Parisi, A. and Petronzio, R., {\it Nucl. Phys.} 
{\bf B158}, 295 (1979).

For more recent improvements see

\noindent$\bullet$ 
Lindner, M., {\it Z. Phys.} {\bf C31}, 295 (1986); Lindner, M., 
Sher, M. and Zaglauer, H.W., {\it Phys. Lett.}{\bf 
B228}, 139 (1989); Sher, M., {\it Phys. Lett.}{\bf B317}, 159 (1993); 
{\it ibid.}{\bf B331}, 448 (1994); Ford, C. et al., {\it Nucl. Phys.}{\bf 
B395}, 17 (1993).

The calculation of the SM Higgs  stability bound including 
next-to-leading log corrections as described in the text is in

\noindent$\bullet$ Casas, J.A., Espinosa, J.R. and Quir\'os, M., {\it Phys. 
Lett.}{\bf B342}, 171 (1995); and Preprint DESY 96-021 [hep-ph/9603227] 
to appear in {\it Phys. Lett.} {\bf B}.

For an independent analysis along similar lines see
 
\noindent$\bullet$ Altarelli, G. and Isidori, I., {\it Phys. Lett.} {\bf B337}, 
141 (1994).

For studies of the renormalization-group improved effective potential
see

\noindent$\bullet$ Coleman, S. and Weinberg, E., {\it Phys. Rev.}{\bf D7}, 1888 
(1973).

\noindent$\bullet$ 
Kastening, B.,  {\it Phys. Lett.}{\bf B283}, 287 (1992); Ford, C. 
et al., {\it Nucl. Phys.}{\bf B395}, 17 (1993); Bando, M., et al., 
{\it Phys. Lett.}{\bf B301}, 83 (1993); {\it 
Prog. Theor. Phys.}{\bf 90}, 405 (1993).
\vskip .2 cm

\noindent
{\it \Ref.4 Section 5}
\vskip .1 cm
For an introduction to the decay of metastable vacuum states see e.g.

\noindent$\bullet$ 
Voloshin, M.B., Lecture at International School of Subnuclear 
Physics, Erice 1995.

Lower metastability bounds on the SM Higgs mass were obtained in:

\noindent$\bullet$ 
Anderson, G., {\it Phys. Lett.}{\bf B243}, 265 (1990); Arnold, 
P. and Vokos, S., {\it Phys. Rev.}{\bf D44}, 3620 (1991).

The next-to-leading log bound is presented in

\noindent$\bullet$ 
Espinosa, J.R. and Quir\'os, M., {\it Phys. Lett.}{\bf B35}, 257 (1995).
\vskip .2 cm

\noindent
{\it \Ref.5 Section 6}
\vskip .1 cm
The interplay between the SM lower bound on the Higgs mass and the upper 
bound for the lightest Higgs boson in the MSSM was considered by

\noindent$\bullet$ 
Krasnikov, N.V. and Pokorski, S., {\it Phys. Lett.}{\bf B288}, 
184 (1992); D\'{\i}az, M.A., ter Veldhuis, T.A. and Weiler, T.J., {\it 
Phys. Rev. Lett.}{\bf 74}, 2876 (1995).

The analysis with the most refined bounds is presented in

\noindent$\bullet$ Casas, J.A., Espinosa, J.R. and Quir\'os, M., {\it Phys. 
Lett.}{\bf B342}, 171 (1995).
\vfill
 \break \bye